\documentclass[twocolumn]{aastex63}
\usepackage{multirow}
\usepackage[T1]{fontenc}
\usepackage{mhchem}
\usepackage{url}
\usepackage{lineno}

\newcommand{\Htwo}{\(\textrm{H}_{2}\)}
\newcommand{\CO}{\(^{12}\textrm{CO}\)}
\newcommand{\thCO}{\(^{13}\textrm{CO}\)}
\newcommand{\CetO}{\(\textrm{C}^{18}\textrm{O}\)}
\newcommand{\CsvO}{\(\textrm{C}^{17}\textrm{O}\)}

\newcommand{\Msun}{\(\textrm{M}_{\odot}\)}

\submitjournal{AJS}

\begin{document}

\title{Molecules with ALMA at Planet-forming Scales (MAPS) XVII: Determining the 2D Thermal Structure of the HD 163296 Disk}

\shorttitle{MAPS XVII: 2D Thermal Structure of the HD 163296 Disk}
%\shortauthors{}

\correspondingauthor{Jenny Calahan}
\email{jcalahan@umich.edu}

\author[0000-0002-0786-7307]{Jenny K. Calahan}
\affiliation{University of Michigan, 323 West Hall, 1085 South University Avenue, Ann Arbor, MI 48109, USA }

\author[0000-0003-4179-6394]{Edwin A. Bergin}
\affiliation{University of Michigan, 323 West Hall, 1085 South University Avenue, Ann Arbor, MI 48109, USA }

\author[0000-0002-0661-7517]{Ke Zhang}
\altaffiliation{NASA Hubble Fellow}
\affiliation{Department of Astronomy, University of Wisconsin-Madison, 
475 N Charter St, Madison, WI 53706}
\affiliation{University of Michigan, 323 West Hall, 1085 South University Avenue, Ann Arbor, MI 48109, USA }

\author[0000-0002-6429-9457]{Kamber R. Schwarz}
\altaffiliation{NASA Hubble Fellowship Program Sagan Fellow}
\affiliation{Lunar and Planetary Laboratory, University of Arizona, 1629 E. University Blvd, Tucson, AZ 85721, USA}

\author[0000-0001-8798-1347]{Karin I. \"Oberg} 
\affiliation{Center for Astrophysics \textbar Harvard \& Smithsonian, 60 Garden St., Cambridge, MA 02138, USA}

\author[0000-0003-4784-3040]{Viviana V. Guzm\'an}
\affiliation{Instituto de Astrofísica, Ponticia Universidad Católica de Chile, Av. Vicuña Mackenna 4860, 7820436 Macul, Santiago, Chile}

\author[0000-0001-6078-786X]{Catherine Walsh}\affiliation{School of Physics and Astronomy, University of Leeds, Leeds, UK, LS2 9JT}

\author[0000-0003-3283-6884]{Yuri Aikawa}
\affiliation{Department of Astronomy, Graduate School of Science, The University of Tokyo, Tokyo 113-0033, Japan}

\author[0000-0002-2692-7862]{Felipe Alarc\'on }
\affiliation{Department of Astronomy, University of Michigan,
323 West Hall, 1085 S. University Avenue,
Ann Arbor, MI 48109, USA}

\author[0000-0003-2253-2270]{Sean M. Andrews} \affiliation{Center for Astrophysics \textbar Harvard \& Smithsonian, 60 Garden St., Cambridge, MA 02138, USA}

\author[0000-0001-7258-770X]{Jaehan Bae}
\altaffiliation{NASA Hubble Fellowship Program Sagan Fellow}
\affil{Earth and Planets Laboratory, Carnegie Institution for Science, 5241 Broad Branch Road NW, Washington, DC 20015, USA}
\affiliation{Department of Astronomy, University of Florida, Gainesville, FL 32611, USA}

\author[0000-0002-8716-0482]{Jennifer B. Bergner} \affiliation{University of Chicago Department of the Geophysical Sciences, Chicago, IL 60637, USA}
\altaffiliation{NASA Hubble Fellowship Program Sagan Fellow}

\author[0000-0003-2014-2121]{Alice S. Booth} \affiliation{Leiden Observatory, Leiden University, 2300 RA Leiden, the Netherlands}
\affiliation{School of Physics and Astronomy, University of Leeds, Leeds, UK, LS2 9JT}

\author[0000-0003-4001-3589]{Arthur D. Bosman}
\affiliation{University of Michigan, 323 West Hall, 1085 South University Avenue, Ann Arbor, MI 48109, USA }

\author[0000-0002-2700-9676]{Gianni Cataldi}
\affil{National Astronomical Observatory of Japan, Osawa 2-21-1, Mitaka, Tokyo 181-8588, Japan}
\affil{Department of Astronomy, Graduate School of Science, The University of Tokyo, Tokyo 113-0033, Japan}

\author[0000-0002-1483-8811]{Ian Czekala}
\altaffiliation{NASA Hubble Fellowship Program Sagan Fellow}
\affiliation{Department of Astronomy and Astrophysics, 525 Davey Laboratory, The Pennsylvania State University, University Park, PA 16802, USA}
\affiliation{Center for Exoplanets and Habitable Worlds, 525 Davey Laboratory, The Pennsylvania State University, University Park, PA 16802, USA}
\affiliation{Center for Astrostatistics, 525 Davey Laboratory, The Pennsylvania State University, University Park, PA 16802, USA}
\affiliation{Institute for Computational \& Data Sciences, The Pennsylvania State University, University Park, PA 16802, USA}
\affiliation{Department of Astronomy, 501 Campbell Hall, University of California, Berkeley, CA 94720-3411, USA}

\author[0000-0001-6947-6072]{Jane Huang}
\altaffiliation{NASA Hubble Fellowship Program Sagan Fellow}
\affiliation{University of Michigan, 323 West Hall, 1085 South University Avenue, Ann Arbor, MI 48109, USA }
\affiliation{Center for Astrophysics \textbar Harvard \& Smithsonian, 60 Garden St., Cambridge, MA 02138, USA}

\author[0000-0003-1008-1142]{John~D.~Ilee} \affil{School of Physics \& Astronomy, University of Leeds, Leeds LS2 9JT, UK}

\author[0000-0003-1413-1776]{Charles J. Law}\affiliation{Center for Astrophysics \textbar Harvard \& Smithsonian, 60 Garden St., Cambridge, MA 02138, USA}

\author[0000-0003-1837-3772]{Romane Le Gal}\affiliation{Center for Astrophysics \textbar Harvard \& Smithsonian, 60 Garden St., Cambridge, MA 02138, USA}
\affiliation{IRAP, Universit\'{e} de Toulouse, CNRS, CNES, UT3, 31400 Toulouse, France}
\affiliation{IPAG, Universit\'{e} Grenoble Alpes, CNRS, IPAG, F-38000 Grenoble, France}
\affiliation{IRAM, 300 rue de la piscine, F-38406 Saint-Martin d'H\`{e}res, France}

\author[0000-0002-7607-719X]{Feng Long}
\affiliation{Center for Astrophysics \textbar Harvard \& Smithsonian, 60 Garden St., Cambridge, MA 02138, USA}

\author[0000-0002-8932-1219]{Ryan A. Loomis}\affiliation{National Radio Astronomy Observatory, 520 Edgemont Rd., Charlottesville, VA 22903, USA}

\author[0000-0002-1637-7393]{Fran\c cois M\'enard}
\affiliation{Univ. Grenoble Alpes, CNRS, IPAG, F-38000 Grenoble, France}

\author[0000-0002-7058-7682]{Hideko Nomura}\affiliation{Division of Science, National Astronomical Observatory of Japan, Osawa 2-21-1, Mitaka, Tokyo 181-8588, Japan}

\author[0000-0001-8642-1786]{Chunhua Qi} \affiliation{Center for Astrophysics \textbar Harvard \& Smithsonian, 60 Garden St., Cambridge, MA 02138, USA}

\author[0000-0003-1534-5186]{Richard Teague}\affiliation{Center for Astrophysics \textbar Harvard \& Smithsonian, 60 Garden St., Cambridge, MA 02138, USA}

\author[0000-0002-2555-9869]{Merel L.R. van 't Hoff}
\affiliation{University of Michigan, 323 West Hall, 1085 South University Avenue, Ann Arbor, MI 48109, USA }

\author[0000-0003-1526-7587]{David J. Wilner}\affiliation{Center for Astrophysics \textbar Harvard \& Smithsonian, 60 Garden St., Cambridge, MA 02138, USA}

\author[0000-0003-4099-6941]{Yoshihide Yamato} \affiliation{Department of Astronomy, Graduate School of Science, The University of Tokyo, Tokyo 113-0033, Japan}

\begin{abstract}

Understanding the temperature structure of protoplanetary disks is key to interpreting observations, predicting the physical and chemical evolution of the disk, and modeling planet formation processes. In this study, we constrain the two-dimensional thermal structure of the disk around Herbig Ae star HD 163296. Using the thermo-chemical code RAC2D, we derive a thermal structure that reproduces spatially resolved ALMA observations ($\sim$0.12 $\arcsec$ (13 au) - 0.25$\arcsec$ (26 au)) of \CO{} $J$ = 2-1, \thCO{} $J$ = 1-0, 2-1, \CetO{} $J$ = 1-0, 2-1, and \CsvO{} $J$ = 1-0, the HD $J$ = 1-0 flux upper limit, the spectral energy distribution (SED), and continuum morphology. The final model incorporates both a radial depletion of CO motivated by a time scale shorter than typical CO gas-phase chemistry (0.01 Myr) and an enhanced temperature near the surface layer of the the inner disk (z/r $\ge$ 0.21). This model agrees with the majority of the empirically derived temperatures and observed emitting surfaces derived from the $J$ = 2-1 CO observations. We find an upper limit for the disk mass of 0.35 \Msun, using the upper limit of the HD $J$ = 1-0 and $J$ = 2-1 flux. With our final thermal structure, we explore the impact that gaps have on the temperature structure constrained by observations of the resolved gaps. Adding a large gap in the gas and small dust additionally increases gas temperature in the gap by only 5-10\%. This paper is part of the MAPS special issue of the Astrophysical Journal Supplement.

\end{abstract}

\keywords{protoplanetary disk, astrochemistry}


\section{Introduction} \label{sec:intro}
The two-dimensional (radial + vertical) thermal structure of protoplanetary disks has been a long sought after property due to its importance in interpreting of observations and its importance on disk evolution. Disk temperature sets the physical and chemical evolution of disk material, with subsequent implications for planet formation. The translation of observations into fundamental disk properties, namely gas mass, strongly depends on the assumed underlying disk temperature at which the observed molecule emits. The mass tracer HD \citep{Bergin13, McClure16, Bergin17, Trapman17, Kama20} especially requires a well defined disk thermal structure as its \textit{J}=1-0 transition has an E$_{\rm{up}}$ of 128.5 K. This approaches typical temperatures within the warm molecular layer of protoplanetary disks. The temperature throughout the disk regulates physical evolution by setting local sound speeds \citep{Bergin17}, turbulent viscosity \citep{Shakura1973, Penna13}, and vertical flaring of the disk \citep{Kenyon87}, which can in turn set the angle of incidence of stellar irradiation, further changing the thermal structure. From a chemical perspective, gas temperature controls the rate of gas-phase exothermic reactions. Temperature also dictates the rates of gas-phase deposition and thermal sublimation, effectively prescribing the relative radial chemical composition of ices in the midplane. These sublimation fronts, or snowlines, have been theorized to be favorable sites for planet formation \citep{Hayashi81, Stevenson88, Zhang15, schoonenberg17}, and may affect the chemical composition of a planetary embryo or accreting atmosphere \citep[e.g. C/O][]{Oberg11, Oberg16, Cridland20}    

The most robust attempts to uncover the temperature profile of protoplanetary disks involve a convergence of observations of multiple gas emission lines and thermo-chemical modeling \citep[e.g][]{Kama16, Woitke19, Rab20, Calahan20}. With the advent of the Atacama Large (sub-)Millimeter Array (ALMA), spatially resolved observations of disks became feasible, providing the first empirical measurements of the radial distribution of dust and gas at high-angular resolution. This gave much needed constraints for protoplanetary disk models. The commonly observed \CO{} $J$ = 2-1 is optically thick in most gas-rich disks due to it being the second most abundant molecule in the gas phase after \Htwo{}. Lower energy transitions and less abundant isotopologues emit from lower vertical layers, resulting in intensity profiles that are affected by temperature and CO surface density/abundance at which that species is emitting. Thus, joint modeling of multiple spatially resolved CO isotopologues provides radial and vertical constraints on the temperature and CO chemistry.

\begin{deluxetable*}{cccccc}
    \label{tab:obs_params}
    \tablecolumns{7} 
    \tablewidth{0pt}
    \tablecaption{ALMA Observations Summary}
     \tablehead{Transition& Frequency & E$_{\rm up}$ & Beam & PA & rms\\
       & (GHz) & (K)& ($\arcsec\times\arcsec$)& ($^\circ$) &(mJy beam$^{-1}$)}
    \startdata
    \CO{} $J$=2-1 & 230.538 & 16.6 & $0.14\times0.10$ & $-$76.3 & 0.561  \\
    \thCO{} $J$=1-0 & 110.201  & 5.3& $0.28\times0.22$ & $-$89.1 &0.434\\
    \thCO{} $J$=2-1 & 220.399 & 15.8 &$0.14\times0.11$ & $-$76.6 & 0.541\\
    \CetO{} $J$=1-0 & 109.782 & 5.3 &$0.28\times0.22$ & $-$88.8 & 0.449\\
    \CetO{} $J$=2-1 & 219.560 & 15.8 &$0.14\times0.11$ & $-$76.5 &0.385\\
    \CsvO{} $J$=1-0 & 112.359  & 5.4&$0.28\times0.21$ & $-$89.5 & 0.528\\
    \enddata
    \tablecomments{These data were taken from values in \citet{Oberg20} and \citet{Czekala21}, where details regarding the data reduction can also be found. }
    \label{tab:obs_params}
\end{deluxetable*}

This study determines the 2D temperature structure of the disk around HD 163296, as part of the ALMA-MAPS project, which observed five protoplanetary disks at high resolution. Each disk is detailed in \citet{Oberg20}. HD 163296  is a Herbig Ae star with a stellar mass of 2.0 \Msun \citep{Andrews18,Wichittanakom20} surrounded by a massive disk ($8\times 10^{-3}$ - $5.8\times 10^{-1}$ \Msun) \citep[][and references therein]{Kama20}  101 pc away \citep{Gaia18, Bailer-Jones18}. Both millimeter continuum and scattered light observations show significant substructure \citep{Grady00, Wisniewski08, Benisty10, Garufi14, Monnier17, Muro-Arena18, Isella18, Dent19, Rich20} including three gaps in millimeter continuum observations at 10, 48, and 86 au. These gaps have been prime targets to test planet formation theories; planets ranging in mass 0.07-4.45 M$_{\rm{J}}$ have been theorized to exist within the gaps in HD 163296 \citep{Zhang18}. The temperature, both of the gas and dust populations in the gaps can be used to directly inform planet formation models. Kinematic studies of HD 163296 suggest the existence of a $\sim$ 2 M$_{J}$ planet at 260 au \citep{Pinte18,Pinte20}. Further analysis of the velocity structure by \citet{Teague19} mapped out the 3D gradient of velocity using \CO{} $J$=2-1 from the DSHARP project \citep{Andrews18}, and in \citet{Teague20}. Meridional flows are found at the location of the continuum gaps, indicative of ongoing planet formation. The protoplanetary disk around HD 163296 presents an excellent case for determining the 2D thermal profile using some of the highest resolution data available for a disk in which planets are believed to be actively forming. 
 
Two-dimensional temperature structures for HD 163296 have been modeled previously using a combination of spatially unresolved observations of the CO rotational ladder, and other higher energy molecular and atomic transitions \citep{Qi11, Rosenfeld13, deGregorio13} with a few studies utilizing thermo-chemical modeling to match numerous observations \citep{Tilling12, Fedele16, Woitke19}. These studies found that HD 163296 is a relatively cooler disk compared to other disks surrounding Herbig stars. Resolved observations of \CO{} $J$=2-1 have also revealed insight into the temperature within the disk gaps \citep{Rab20}. 
%with results ranging from a very flat gaseous disk \citep[$\sim$ 0.05][]{Tilling12,Kama20} to a flared disk surface \citep[0.25-1.08][]{Rosenfeld13, Fedele16,Isella16,Rab20}. 

The observations from the ALMA-MAPS program consists of highly resolved ($\sim$ 0.12$\arcsec$ - 0.25$\arcsec$), and high S/N observations of \CO{} $J$ = 2-1, \thCO{} $J$ = 2-1, 1-0, \CetO{} $J$ = 2-1, 1-0, and \CsvO{} $J$ = 1-0. These observations show structure in the radial intensity profiles \citep{law20rad} and CO column density \citep{Zhang20b} (hereafter Z21) and allow for a new derivation of the 2D thermal structure. % An accurate high resolution thermal structure also allows for insight into the temperature structure within gas gaps.

This paper is organized as follows: We detail our modeling procedure and describe the observations in Section \ref{sec:method}. In Section \ref{sec:results}, we present our best fit thermal model and the necessary alterations needed to fit each observed line. In Section \ref{sec:diss}, comparisons of the final model to the empirically derived temperature structure and emitting surfaces are discussed along our derivation of an upper mass limit for the HD 163296 disk and explore CO/\Htwo{} degeneracy and temperature effects within the two largest gaps of the HD 163296 disk. We summarize our findings in Section \ref{sec:conclusion}.

\section{Methods and Observations}\label{sec:method}
\subsection{Observations}
We use observations of \CO{} $J$=2-1, \CetO{} $J$=2-1, \\1-0, \thCO{} $J$=2-1, 1-0, and \CsvO{} $J$=1-0 towards the HD 163296 disk. For this study, we use images which have an average spatial resolution of 26 au for the 1-0 transitions (0.25$\arcsec$) and 13 au for the 2-1 (0.12$\arcsec$) transitions (\citet{Oberg20}, and Table \ref{tab:obs_params}). The final image cubes used for this study had a spectral resolution of 200 m s$^{-1}$, and used a robust = 0.5 weighting. The robust = 0.5 images were used as they provided the highest absolute resolution. These images were ``JvM-corrected'' \citep{Jorsater95}, which refers to the method of scaling the residuals in the image cube to have identical units as the CLEAN model. This ensures that the final image used for moment zero maps, and subsequent radial profiles, have the correct units (Jy $\mathrm{CLEAN\,beam}^{-1}$); see \citet{Czekala21} for a detailed explanation. The MAPS program summary, along with data reduction and calibration specifics can be found in \citet{Oberg20}, and the methods of the imaging process are throughly described in \citet{Czekala21}. Radial intensity profiles are created for each CO line for comparison to our model results. Radial profile derivations are presented in \citet{law20rad}. The package  \textsc{bettermoments}\footnote{https://github.com/richteague/bettermoments} \citep{Teague18} is used to extract the moment zero map which is then used to calculate the radial intensity profile using \textsc{GoFish}\footnote{https://github.com/richteague/gofish} \citep{gofish}. We follow the same methods in creating our simulated radial intensity profiles.

In addition to ALMA observations of CO, we also compared our model to the \textit{Herschel Space Observatory} observation of HD $J$= 1-0 and 2-1 towards HD 163296. The HD observations were made using the PACS instrument and were spatially and spectral unresolved. \citet{Kama20} used archival data from the DIGIT program (PI N.J. Evans) to determine the upper flux limits for 15 Herbig Ae/Be disks, including HD 163296. They find an upper limit for the flux of the $J$= 1-0 line of $\leq$~0.9~$\times$~10$^{-17}~\rm{W}/\rm{m^{2}}$ and for the $J$= 2-1  they determine a flux of $\leq$~2.4~$\times$~10$^{-17}~\rm{W}/\rm{m^{2}}$. 

\subsection{RAC2D Physical Structure}

In this study, we use the time-dependent thermo-chemical code RAC2D\footnote{https://github.com/fjdu/rac-2d} \citep{Du14} to model the thermal structure of the HD 163296 disk, and create simulated observations to compare to the ALMA data. A brief description of the physical code of RAC2D is given below; a detailed description of the code can be found in the aforementioned paper. 

RAC2D takes into account both the disk gas and dust structure while simultaneously computing the temperature and chemical structure over time. Our model consists of three mass components: gas, small dust  (\(\leq\) \micron), and large dust grains (\(\leq\) mm). The distribution of each component is described by a global surface density distribution \citep{Lynden-Bell74}, which is widely used in modeling of protoplanetary disks and corresponds to the self-similar solution of a viscously evolved disk.

\begin{equation}
\Sigma(r)=\Sigma_{c}\left(\frac{r}{r_{c}}\right)^{-\gamma}\exp{\left[-\left(\frac{r}{r_{c}}\right)^{2-\gamma}\right]},
\label{surfden_equn}
\end{equation}

\noindent \(r_{c}\) is the characteristic radius at which the surface density is \(\Sigma_{c}/\rm{e}\) and \(\gamma\) is the power-law index that describes the radial behavior of the surface density.

A density profile for the gas and dust components can be derived from the surface density profile and a scale height:

\begin{equation}
    \rho(r,z) = \frac{\Sigma(r)}{\sqrt{2\pi}h(r)} \exp{\left[-\frac{1}{2}\left(\frac{z}{h(r)}\right)^{2}\right]},
\label{density_equn}
\end{equation}

\begin{equation}    
    h=h_{c}\left( \frac{r}{r_{c}}\right)^{\Psi} ,
\label{heigh_equn}
\end{equation}

\noindent where \(h_{c}\) is the scale height at the characteristic radius, and \(\Psi\) is a power index that characterizes the flaring of the disk structure.

Both dust populations follow an MRN grain distribution \(n(a) \propto a^{-3.5}\) \citep{Mathis77}. The small dust grains have radii between $5 \times 10^{-3}$ - $1 \mu$m, and the large grains have radii between $5 \times 10^{-3}$ - $10^{3} \mu$m. The following description of the dust grain population used in this study is adopted from Z21, which used both the HD 163296 SED and mm continuum observations to constrain the dust population. The gas and small dust grains are spatially coupled and extend to 600 au. The large dust population is settled in the midplane with a smaller vertical extent ($h$ = 1.69~au) and radial extent (240~au). This settled large grain population is the result of dust evolution, namely vertical settling to the midplane and radial drift. The large grain population has a unique, non-smooth, surface density profile that reproduces the millimeter continuum observations of the HD 163296 disk (Z21). The dust composition adopted is consistent across the MAPS collaboration, and opacity values are calculated based on \citet{Birnstiel18}. Large dust grains consist of water ice \citep{Warren08}, silicates \citep{Draine03}, troilites and refractory organics \citep{Henning96}. Small dust grains consist of 50\% silicates and 50\% refractory organics.    

\subsection{RAC2D Dust and Gas Temperature}
The code computes a dust and gas temperature after initializing RAC2D with a model density structure for each population, a stellar spectrum, and external UV radiation field (G$_{0}$=1). The stellar spectrum we use is a composite of multiple UV and X-ray observations, and is further detailed in Z21. The determination of dust and gas temperature is an iterative process, allowed to change over time due to the evolving chemical composition. The dust thermal structure is calculated first using a Monte Carlo method to simulate photo absorption and scattering events. This also results in a radiation field spanning cm to X-ray wavelengths. 

We adopt the reaction rates from the UMIST 2006 database \citep{Woodall07} for the gas-phase chemistry with additional rates considering the self-shielding of CO, \Htwo{}, \(\textrm{H}_{2}\textrm{O}\), and OH, dust grain surface chemistry driven by temperature, UV, cosmic rays, and two-body chemical reactions on dust grain surfaces \citep[see references given by][]{Du14}. Chemical processes also provide heating or cooling to the surrounding gas. These mechanisms along with stellar and interstellar radiation drive the thermal gas structure. Our study explores models that account for 0.01 Myr of chemical evolution. This is a relatively short period of time compared to disk lifetimes and is motivated by an average CO processing timescale, which is found to be on the order of 1~Myr \citep{Bergin14,Bosman18,Schwarz18_1}.
%which is motivated by the fact that we deplete the abundance of CO before the temperature and chemical calculations. 
The calculated depletion profile of CO is motivated by observations, and thus accounts for any long time-scale chemical effects that have taken place in the disk's history; we did not want to further process CO by running the disk model for longer than 0.01 Myr. It is also worth noting that the exact time scale will not affect our final temperature structure nor CO column density distribution; it will only affect the CO depletion profile. CO will be created or destroyed in our chemical network over time, and the CO depletion profile acts to counter any over or under-abundance of CO being produced. While localized variations in the CO abundance may affect the gas temperature, the global temperature structure is not strongly affected by the CO abundance.%This timescale is sufficiently long and allows for the gas temperature to converge to the dust temperature for the majority of the disk, including the regions of the disk in which where our observations emit. 
Finally, at the end of a given run, we extracted the dust and gas thermal profiles. 

Simulated line images for CO, \thCO{}, \CetO{}, \CsvO{} and HD are necessary for our comparison to observations. We did not model isotopologue fractionation in this chemical network, as fractionation of CO is not significant in a massive disk like HD 163296 \citep{Miotello14}. Thus, we computed \thCO{} and \CetO{} abundances based on ISM ratios of \CO{}/\thCO{} = 69, \CO{}/\CetO{} = 557, \CetO{}/\CsvO{} = 3.6 \citep{Wilson99}.  Given these abundances and the local gas temperature, RAC2D computes synthetic channel maps using a ray-tracing technique. We then convolved these simulated observations with the corresponding ALMA CLEAN beam to make direct comparisons to data. To directly compare to observations, we created simulated integrated radial intensity profiles of each of the CO lines, following the methods from \citet{law20rad}. Our simulated observations have the same beam sizes, spectral resolution, and pixel size as what is reported in \citet{Oberg20}. To recreate the unresolved integrated flux measurements for HD, we convolved our model HD lines with a Gaussian profile corresponding to the velocity resolution of the \textit{Herschel} PACS instrument: $\sim$300 ${\rm km\,s^{-1}}$ at $\sim$112 $\mu$m and $\sim$100 ${\rm km\,s^{-1}}$ at $\sim$51 $\mu$m \citep{Poglitsch10}

\begin{deluxetable}{cc}
    \tablewidth{0pt}
    \tablecolumns{2}
    \tablecaption{Initial Chemical Abundances for Final Model}
    \tablehead{
    &\colhead{Abundance Relative to Total}\\
    &\colhead{Hydrogen Nuclei}}
     \startdata
      \Htwo & \( 5 \times 10^{-1}\) \\
      HD & \(2 \times 10^{-5}\) \\
      He & 0.09\\
      CO$^{a}$ & see Figure \ref{COdep}\\
      N & \(7.5 \times 10^{-5}\)\\
      \(\textrm{H}_{2}\textrm{O (ice)}\) & \(1.8 \times 10^{-4}\)\\
      S & \(8 \times 10^{-8}\)\\
      Si$^+$ & \(8 \times 10^{-9}\)\\
      Na$^+$  & \(2 \times 10^{-8}\)\\
      Mg$^+$  & \(7 \times 10^{-9}\)\\
      Fe$^+$  & \(3 \times 10^{-9}\)\\
      P & \(3 \times 10^{-9}\)\\
      F & \(2 \times 10^{-8}\)\\
      Cl & \(4 \times 10^{-9}\)\\
     \enddata
     \tablecomments{$^{a}$CO starts with an abundance of \(1.4 \times 10^{-4}\) with an} imposed radial depletion profile as shown in Figure \ref{COdep}
     \label{chemAbundance}
\end{deluxetable}

\begin{figure}
    \centering
    \includegraphics[scale=.45]{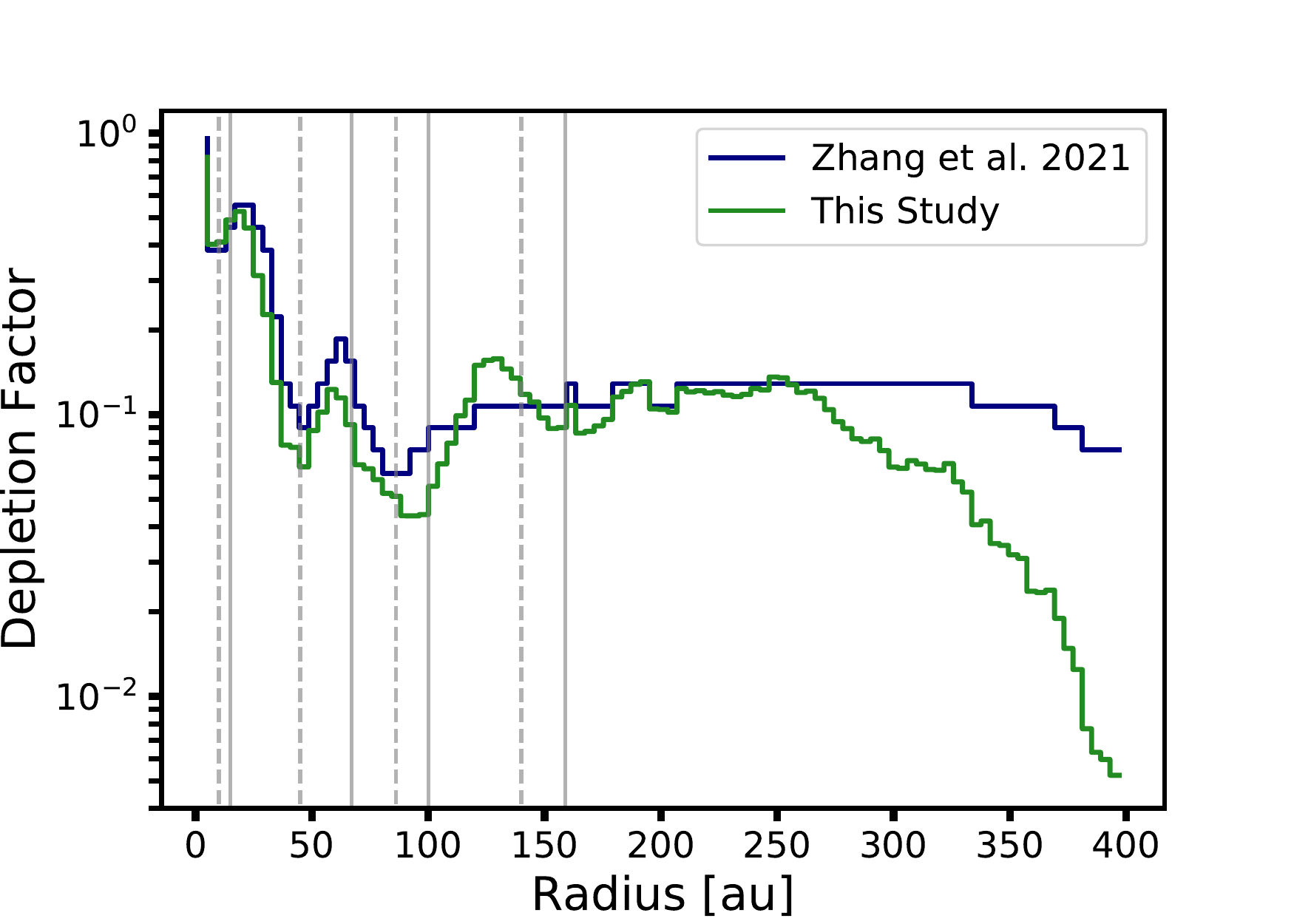}
    \caption{Multiplicative depletion factor for the initial CO abundance (CO/H=\(1.4 \times 10^{-4}\)) as shown in Table \ref{chemAbundance}. Light gray lines indicate the location of the dust rings (solid) and gaps (dashed). The line in blue shows the CO depletion as derived by Z21 and is determined after an initial thermo-chemical model of the HD 163296 disk and is based on the \CetO{} $J$=2-1 observation, and acts as a starting point for our CO depletion determination. The green represents the final CO depletion, also motivated by the \CetO{} $J$=2-1 flux but is taken into account at the beginning of the chemical and thermal calculations.}
    \label{COdep}
\end{figure}

We began with a model of the HD 163296 disk from Z21, which starts with an initial set of disk parameters taken from the literature \citep[see references within][] {Zhang20b} and applied the observed UV, X-ray and photosphere  stellar spectra. These authors then determined a gas and dust density and dust temperature structure by matching the SED and ALMA continuum image using RADMC3D \citep{radmc3d}. Given this initial density and dust temperature distribution, RAC2D is then used to compute the gas temperature and disk chemistry. Z21 found that in order for the simulated radial profiles to match what is observed, they must modify the CO abundance relative to \Htwo{} as a function of radius. They use the difference between \CetO{} $J$=2-1 as observed and as simulated to predict a CO depletion profile. That depletion profile is shown in Figure \ref{COdep}. The initial chemical abundances used in Z21 are shown in Table \ref{chemAbundance} and final model parameters are summarized in Table \ref{params}. In the study described here, we use the Z21 model as a starting point to derive a 2D temperature map incorporating additional constraints from CO isotopologues, multiple higher-level CO transitions and HD flux. We note that while any revisions made were important to constrain the disk temperature structure, they do not affect the results reported in Z21, and the final CO column densities from this model agree well with what is found in  Z21.

\begin{deluxetable}{cccc}
    \tablewidth{0pt}
    \tablecolumns{2}
    \tablecaption{Gas and Dust Population Parameters: Best-fit Model Values}
    \tablehead{
    &Gas& Small Dust  & Large Dust}
     \startdata
     Mass (\Msun) & 0.14 & \( 2.6 \times 10^{-4}\) & \( 2.4 \times 10^{-3} \) \\
     \(\Psi\) & 1.08 & 1.08  & 1.08 \\
     \(\gamma\) & 0.8  & 0.8  & 0.1 \\
     \({\rm h}_{\rm c}\) [au] & 14.5 &14.5 &  $-$ \\
     \({\rm r}_{\rm c}\) [au] & 165 & 165 &  $-$ \\
     \(r_{\rm in}\) [au] & 0.45 & 0.45 & 0.45 \\
     \(r_{\rm out}\) [au] & 600 & 600 & 240  \\
     \enddata
     \tablecomments{Final values of the HD 163296 model that reproduces the CO, HD, and SED observations. Small dust grain sizes range from \(5 \times 10^{-3}\) - 1\(\micron\), large dust grain sizes range from \(5\times 10^{-3}\) - \(1 \times 10^{3}\)\(\micron\). The large dust population does not have associated \({\rm h}_{\rm c}\) nor \({\rm r}_{\rm c}\) because the surface density is set by continuum observations, thus is non-smooth and not dictated by Equation \ref{surfden_equn}-\ref{heigh_equn}}
     \label{params}
\end{deluxetable}

\section{Results}\label{sec:results}

\begin{figure*}
    \centering
    \includegraphics[scale=.5]{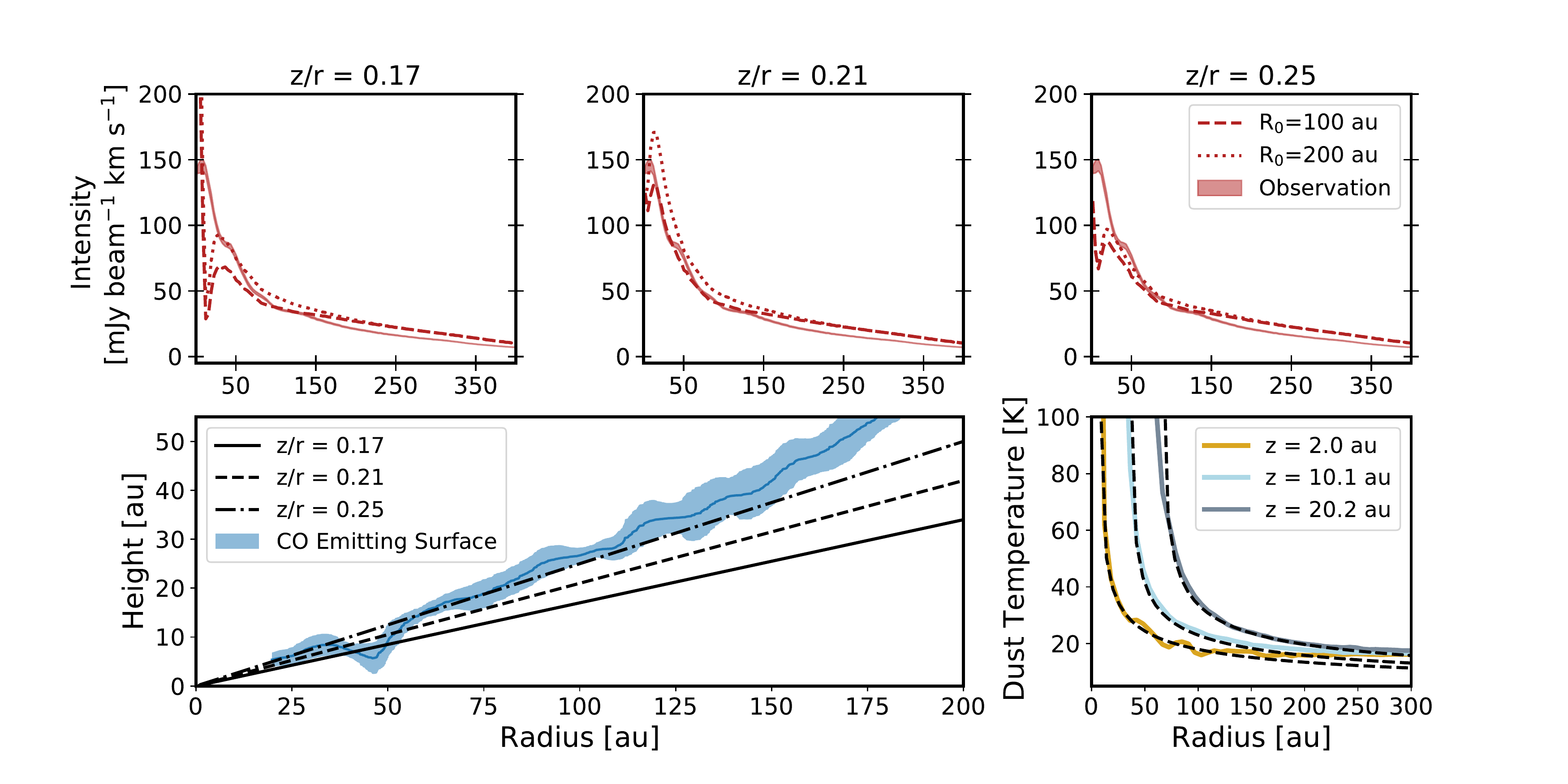}
    \caption{These figures represent the effect of excess heating on the \CO{} $J$=2-1 line profile given different height over radius (z/r) limits (0.17, 0.21, 0.25, columns) and R values (100 and 200 au, dashed line and dotted line) using $(r/R{_{0}})^{-0.4}$ to calculate the excess factor of heat in the regions above a certain z/r. We sought to find a z/r and R$_{0}$ value that would reproduce the observed \CO{} $J$ = 2-1 radial profile (solid red lines). The above is just a sample of the z/r and R$_{0}$ values explored, with our final model using z/r limit of 0.21 and R$_{0}$ = 100 au. The bottom left plot shows the observed emitting surface of \CO{} $J$ = 2-1 (blue) and the z/r values explored. The bottom right plot shows the dust temperature over radius at different heights in the disk, which follow an inverse power-law function, and each dashed black line is a power-law following r$^{-0.4}$, which is what we use to determine the excess heat factor. }
    \label{fig:excessheat_explore}
\end{figure*}

\subsection{CO depletion}\label{sec:COdep}
%To match all six CO radial profiles, the CO depletion profile derived from MAPScoco model had to be altered.

In Z21, depletion of CO was calculated based on a model of the disk based on previous determinations of disk parameters and the \CetO{} $J$ = 2-1 radial intensity profile. The CO was depleted throughout the disk at the start of the raytracing method which creates the simulated observations, i.e. after the temperature and chemical evolution of the disk. This depletion means the CO abundance is reduced from its expected value ($\sim 10^{-4}$ relative to \Htwo{}) in layers where the dust temperature is above the CO sublimation point and the gas is self-shielded from radiation. Since CO is a significant coolant in the disk surface layers where the dust and gas are thermally decoupled, any depletion of CO may affect the thermal structure. Thus, for this study, we recalculated the CO depletion factors. We started by applying the Z21 CO depletion to the initial CO abundance. After running the model for 0.01 Myr, there were small inconsistencies between the simulated and observed \CetO{} $J$=2-1 radial emission profiles (see Figure \ref{fig:init}). We then calculated a new CO depletion profile by determining what factor of increase or decrease in CO flux would be needed to reproduce the observation at each radii (using the same radial resolution as Z21). That factor was the applied to the original CO depletion profile at the corresponding radius. The model was then run again with the new CO depletion profile. We needed to iterate this process three times, and stopped iterating once the $\chi^{2}$ value (using \texttt{stat.chisquare} function from the \texttt{scipy} package) comparing the simulated and observed \CetO{} $J$=2-1 radial profile intensity values had dropped well below that of the first attempts. The first three attempts had $\chi^{2}$ value of 8.06-9.72, and the final model $\chi^{2}$ = 2.47. The final depletion profile is shown in Figure \ref{COdep}, and on average the newly calculated CO depletion is 33\% more depleted than that derived in Z21. Most of the difference is at large radii where the profiles vary by only 16\% within 200 au. 

There is a strong decrease in CO abundance beyond $\sim$250 au, which is not seen in the CO depletion profile calculated by Z21. The CO depletion profile from Z21 accounts for an initial CO depletion in HD 163296 plus 1~Myr of CO processing accounted for in the RAC2D chemical evolution. In our model, we include only 0.01~Myr of additional chemical processing, thus we do not attempt to model the full evolution of CO. In the Z21 model, their derived CO depletion profile takes into account chemical processing that occurs over the course of 1~Myr. The difference in CO abundance for our HD 163296 disk model as compared to the Zhang model is then attributed to chemical CO depletion mechanisms that occur over 1~Myr. In the Zhang model, fully self-shielded CO becomes frozen out and interacts with OH frozen out onto grains and creates CO$_{2}$. In our model, we treat past CO chemistry as an initial condition. We find it necessary to additionally deplete CO  beyond $\sim$240 au in the model that runs for only 0.01~Myr, since the pathway to convert CO and CO$_{2}$ is unavailable within that time.

%0.01~Myr of chemistry, there is not enough OH frozen onto grains for the creation of CO$_{2}$. The main OH formation pathway is through UV processing of \HtwoO{} into OH. In 1~Myr there is enough time to produce OH, especially beyond the large dust population, which is the main opacity source for UV rays and ends at 240 au.  Thus, we find it necessary to additionally deplete CO  beyond $\sim$240 au in the model that runs for only 0.01~Myr, since the pathway to convert CO and CO$_{2}$ is unavailable within that time. 

%in the outer disk, which could be attributed to the low density environment allowing for more UV penetration and results in the destruction of CO \citep{Schwarz18, Bosman18}.

%However it reaches such low levels of CO that are difficult to reproduce by only chemical processes \citep{Schwarz18}. This feature could also be attributed to gas dissipation and dispersal from the outer regions of the disk. Winds have been seen in kinematic data from \CO{} observations in both in the DSHARP data set \citep{Teague19} and this MAPS dataset \citep{Teague20}, which could cause gas dispersal out of the disk and would be reflected in the CO depletion profile due to the degeneracy between CO and \Htwo{} surface densities \citep{Calahan20}.

\subsection{Additional Heating}\label{sec:heat}

While our initial model with the updated CO depletion profile reproduced most of the observed CO lines, \CO{}  $J$=2-1 was under-predicted on average by a factor of 1.6 within 100 au, and a factor of 2.2 within 35 au. We completed a thorough exploration of the parameter space including gas mass, small dust mass, surface density power-law index ($\gamma$), flaring index ($\Psi$), scale height (h$_{\rm{c}}$), critical radius (r$_{\rm{c}}$), and outer radius (see Appendix Section \ref{sec:param}). However, we found no combination of parameters that could improve the \CO{}  $J$=2-1 flux agreement while leaving the other lines consistent with observations. This result together with the high optical depth of the \CO{} $J$ = 2-1 transition suggests that the discrepancy is due to a higher temperature in the CO emitting layer. This has also been seen in \citet{deGregorio13}; \CO{} $J$ = 3-2 was observed to be brighter than their model which matches the outer disk. To solve this issue they increased the gas temperature beyond the dust temperature within 80 au in their HD 163296 disk model.

This higher gas temperature requires additional heating, beyond the level generated by the UV and X-ray field in our model. We began by isolating the layers in which the gas and dust thermal structure are decoupled. Within those regions, with the goal of representing excess heating due to radiation from the star, we artificially increased the gas temperature after the thermo-chemical calculation, and simulated new CO observations. We increased the temperature in this region following a power-law dependent on radius, and by a constant amount vertically (see equation below). There was no realistic amount of heating within the decoupled layer that would increase the intensity of the \CO{}  $J$=2-1 line to match what is observed, any excess heating only affected emission from the inner few au. This then suggests that the \CO{}  $J$=2-1 primarily emits from the region lower in the disk, in which the dust and gas temperatures are coupled, and the necessary excess heating would decouple the gas and dust temperatures. 

We increase the gas temperature radially following the dust thermal profile. Increasing the gas temperature as function of radius following a power-law, 

\[T_{\rm{new}} = T_{\rm{original}} \times(\frac{\rm{r}}{100 \rm{au}})^{-0.4}\]

\noindent above a constant ratio of height over radius (z/r) = 0.21 reproduces the observed \CO{}  $J$=2-1 radial profile. Outside of this region, the gas and dust temperatures remain coupled. Various z/r limits were explored, and z/r = 0.21 produced a result that appeared to significantly affect the \CO{}  $J$=2-1 line while not altering any other lines (see Figure \ref{fig:excessheat_explore} for a sample of the parameter exploration results). It can then be presumed that the emitting surface of \CO{}  $J$=2-1 exists at or above z/r=0.21 within 100 au, and all other lines emit from below these heights (this is supported by further analysis probing the emitting surfaces of each 2-1 transition, see Section \ref{sec:emit}, and \citet{law20emit}). Higher transitions of CO, up to \CO{} $J$ = 23-22 have been observed using the SPIRE instrument on \textit{Herschel} and are compiled in \citet{Fedele16}. These transitions will primarily emit from the very inner regions of the disk, and would be affected by the increase in temperature. Our final model reproduces all of the observed fluxes of the 19 higher-level transitions of CO, most within 1$\sigma$ uncertainly, and five of the transitions fit within 2 or 3$\sigma$ uncertainty (see Table \ref{tab:upperlevelCO}). While the flux predictions from the model all fall within 3$\sigma$, all are on the fainter end of the observed flux. Additionally, the HD J=2-1 flux does not change significantly, increasing only by 7\%, remaining well below the observed upper limit.
%Thus, a model without this increase in temperature would more poorly reproduce the upper-level CO observations.
A model without the additional heat produces a CO flux from the $J$=11-10 transition that is over two times less than our final model. On average, for the transitions between $J$=11-10 and $J$=23-22, a model without the additional heat produces a CO flux over four times lower than our current model. 

The majority of the excess heat is added within 25 au where gas temperatures are beyond twice the original output from our thermo-chemical model. A likely source of this excess heat is mechanical heating from processes such as stellar winds or dissipation of gravitational energy from accretion through the disk onto the star \citep[e.g.][]{Glassgold04,Najita17}. Mechanical heating from such phenomena is expected to be prominent in the inner disk ($<$ 10 au), and are not accounted for in RAC2D. Another possible heating source originates from PAHs, as photoelectric heating of small grains \citep{Draine01,Li01} is one of the main heating mechanisms in this region after direct UV heating from the star. The PAH abundance might be enhanced in the inner disk if dust sintering is taken into account \citep{Okuzumi16}. As dust grains pass the sublimation front of their constituent material, PAHs can be released, enhancing the effect of photoelectric heating near which $^{12}$CO 2-1 emits. PAH emission towards HD 163296 has been observed as a part of the \textit{ISO}/SWS atlas \citep{Sloan03} and modeled by \citet{Seok17}. Their best fit model uses a PAH abundance of 8$\times$10$^{-7}$~M$_{\Earth}$, which is 1.5$\times$ the abundance used in our RAC2D model. In our model and assumed physical set-up, the excess heat is necessary in a region where the gas and dust are thermally coupled. Any increase in PAH abundance in our model will not change the temperature at which the majority of \CO{} $J$=2-1 emits. It only affects the temperature in thermally-decoupled layers where dust densities are low and gas and dust collisions do not occur often. However, it remains a possibility given an alternate setup of HD 163296 in which \CO{} $J$=2-1 emits in the thermally decoupled layers.

\begin{figure*}
    \centering
    \includegraphics[scale=.5]{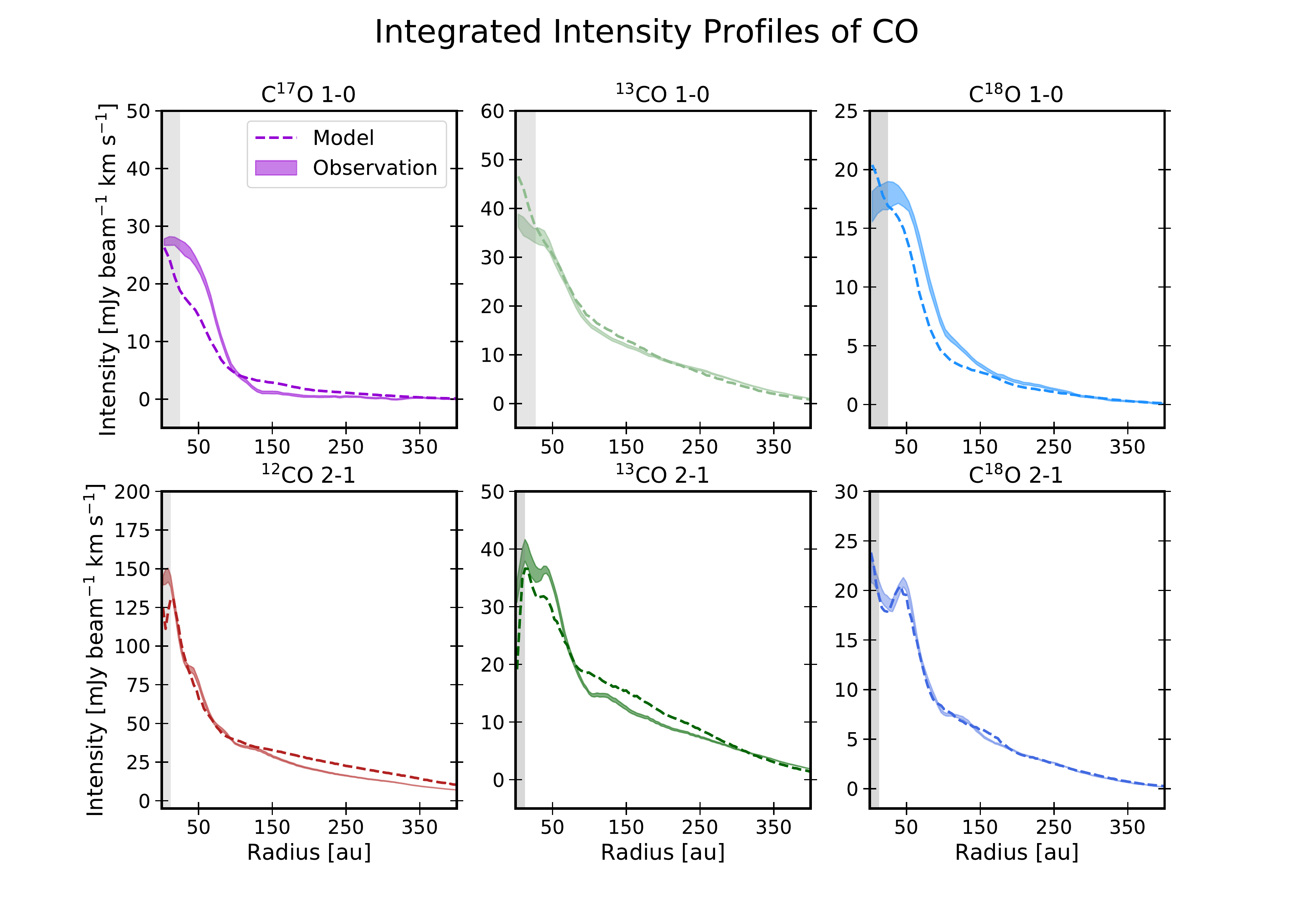}
    \caption{Integrated radial intensity profiles of \CO{} and its isotopologues \thCO{}, \CetO{}, and \CsvO{} as observed (solid line) and as simulated by the thermo-chemcial code RAC2D (dashed line). The gray shaded regions correspond to the FWHM of the corresponding beam for each observed line. This presents our best-fit model to the observations of the HD 163296 disk and represents the thermal structure shown in Figure \ref{temperatures}. }
    \label{COprofs}
\end{figure*}

\begin{figure*}
    \centering
    \includegraphics[scale=.6]{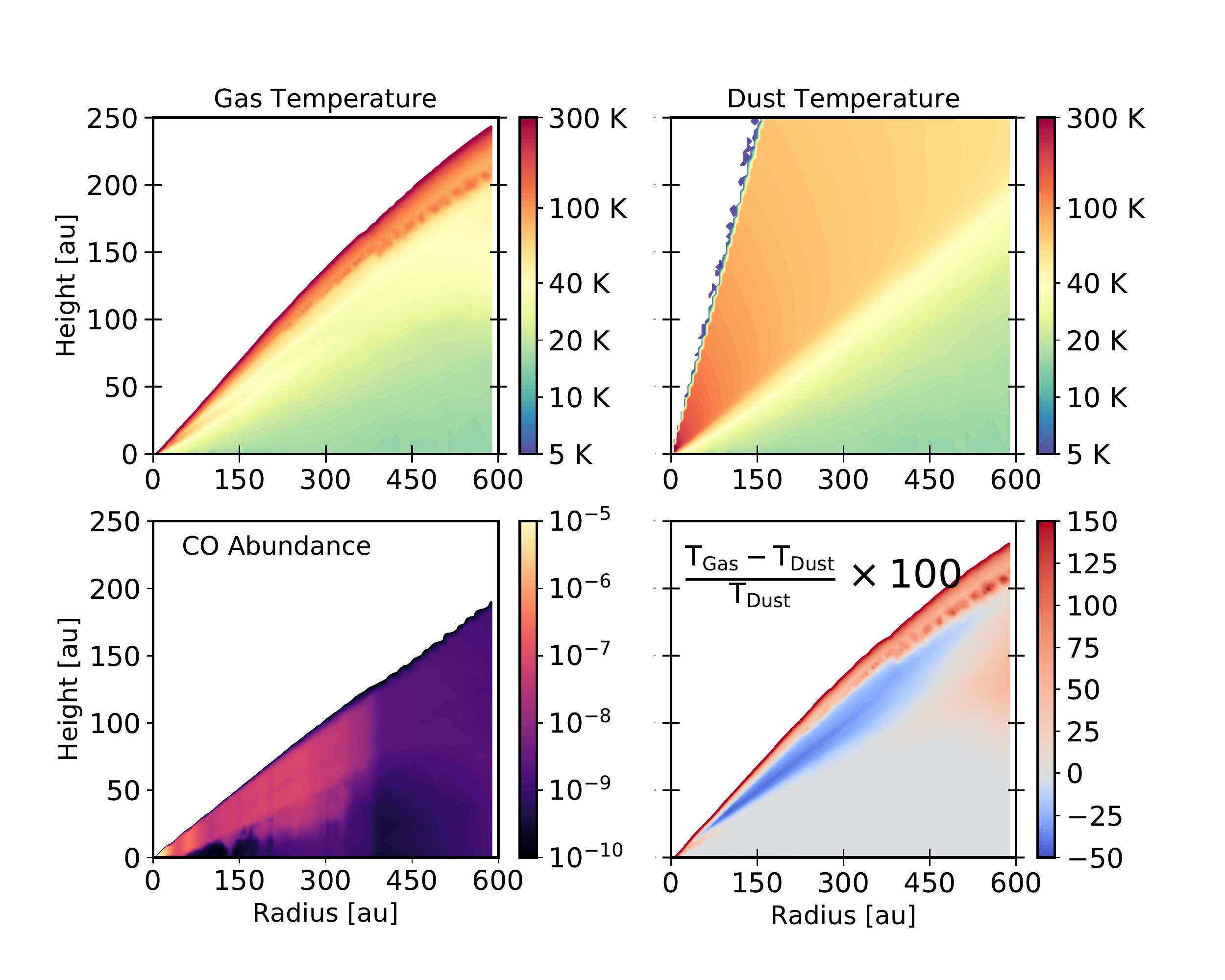}
    \caption{The two-dimensional profile of gas temperature, dust temperature, percentage difference between the two, and CO abundance. The dashed black or white line in each plots shows where z/r=0.21 which corresponds with the \CO{} $J$ = 2-1 surface within 100 au. This represents the model that best-represents the observed radial profile, SED, and an HD flux that agrees with the currently derived upper limit. In the percentage difference plot, one can see where the dust and gas temperatures are coupled (grey), the `undershoot'  region \citep{Kamp04}, where dust is warmer than the gas (blue), and where gas is warmer high up in the atmosphere (red).}
    \label{temperatures}
\end{figure*}

\begin{figure*}
    \centering
    \includegraphics[scale=.43]{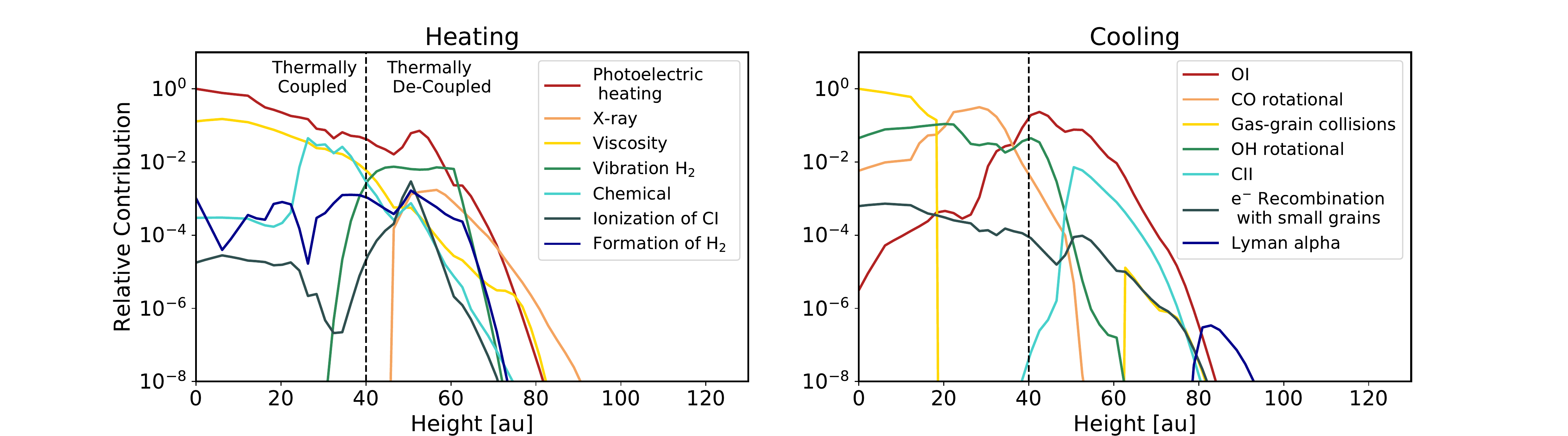}
    \caption{The seven most significant heating (right) and cooling (left) mechanisms in our final model at a disk radius of 150 au, as a function of disk height, normalized according to the highest heating/cooling value at this radius. The black dashed line corresponds to the height (40 au) at which the dust and gas temperatures become decoupled.  }
    \label{fig:heatingandcooling}
\end{figure*}
\subsection{Final Model}\label{sec:model}

The model that best reproduces the CO radial profiles observed towards HD 163296 was achieved by altering the CO depletion profile derived by Z21, and increasing the gas temperature above a z/r = 0.21 within the inner 100 au of the disk. The CO radial profiles derived from the final model are shown in Figure \ref{COprofs} together with the observed profiles. The final gas and dust thermal structures are shown in Figure \ref{temperatures}. Comparing the gas and dust temperatures, we find that below a z/r $\simeq$ 0.25 the dust and gas are thermally coupled (with the exception of the increased gas temperature added artificially). Right above  z/r $\simeq$ 0.25 the dust is hotter than the gas by factors of 25\%-50\%. This region has been referred to as the `undershoot' region \citep{Kamp04} and occurs in disks where the gas density increases sufficiently for line cooling to become very efficient, and occurs around the $\tau = 1$ surface. In our model, atomic oxygen dominates cooling in this region, and cooling via gas-grain collisions are particularly inefficient (see Figure \ref{fig:heatingandcooling}). At the very surface of the disk, the gas temperature then increases drastically, overtaking the dust temperature significantly, which tends to plateau above the undershoot region. Directly above the undershoot region, photoelectric heating and the vibrational transitions of \Htwo{} are the most significant heating processes, with direct heating of the gas via X-ray radiation overtaking beyond z/r $\simeq$ 0.45-0.5. This gas/dust temperature decoupling has been predicted \citep{Kamp04,DAlessio05} and seen in models before \citep{Tilling12, Woitke19, Rab20}.

It is worth noting that the \CsvO{} and \CetO{} 1-0 lines are the least well fit to the observations. We also find that these transitions originate from deep within the disk, only slightly above the midplane, and CO snowsurface. In Z21 it was noted that the \CsvO{} column densities appear to be higher than those estimated from \CetO{} (correcting for optical depth). This is also the case for an earlier $^{13}$C$^{18}$O detection, which appears to require higher CO column than inferred from \CetO{} $J$ = 2-1 (Z21).  A similar analysis is preformed for the MAPS data in Z21. It is suggested that the CO abundance might be higher in the midplane as the icy pebbles drift inwards and sublimate CO inside the snowline. \citet{Booth19} estimates a disk mass of 0.21 \Msun{} using $^{13}$C$^{17}$O $J$=3-2, a value 50\% more massive than our final model. However, if CO is enhanced near the midplane, a larger mass may not be necessary. Such an effect is not accounted for in our analysis, but, at face value, would be consistent with our results in that a locally higher abundance of CO in the midplane, perhaps inside the CO snowline, would increase the emission of optically thin tracers (such as \CsvO{}) while having a reduced effect on the emission of the more abundant isotopologues. This process has negligible effect on the thermal structure as the dust and gas are strongly coupled in these layers.

\subsubsection{CO and CO$_{2}$ Snowline}

To determine the location of the CO snowline in our model, we determine the radial location at which there are equal parts CO frozen onto the dust and in the gas phase. RAC2D includes adsorption of CO (and other species) onto dust grains as well as desorption from the dust surface either thermally, or via UV photons or cosmic rays \citep[see][]{Du14}. Using this metric, our CO snowline is located at 60 au at a temperature of 18 K. This falls between the largest gaps in the continuum, and is consistent with Z21. %which predicts a snowline location of 65 au.
\citet{Qi15} used observations of N$_2$H$^+$ towards HD 163296 to determine the CO snowline as N$_2$H$^+$ formation is inhibited in the presence of gas-phase CO. When CO is frozen-out N$_2$H$^+$ can exist, emitting as a ring with the inner most edge corresponding to the CO snowline. Using this method, and a model of the HD 163296 disk, they found a CO snowline at 74$^{+7}_{-5}$ au (corrected for pre-Gaia distance). Our midplane snowline location is in good agreement with the N$_2$H$^+$ derived snowline location, especially when considering findings from \citet{vantHoff17} that show the N$_2$H$^+$ column density peaks 5 au or more beyond the midplane CO snowline. Additionally we can predict the CO$_{2}$ snowline to be at 4 au, at a temperature of 65 K. These freeze-out temperatures depend on the given desorption energy. RAC2D uses values from \citet{Garrod08} which assumed an amorphous ice surface. Assuming a different desorption energy will alter the derived snowline locations. %This freeze-out temperature for CO$_{2}$ is in agreement with the sublimation temperature of CO$_{2}$ onto pure ices using binding energies found by \citet{Martin-Dominech14}.
It is important to note also, that these snowlines are based on a thermo-chemical model influenced by observations that do not directly probe the midplane. %Our model has a low resolution in the inner disk, thus molecules of a higher freezeout temperature, such as \HtwoO{} are not able to be predicted with this model.

\section{Discussion \& Analysis}\label{sec:diss}

\subsection{Emitting Surface}\label{sec:emit}

\begin{figure}
    \centering
    \includegraphics[scale=.53]{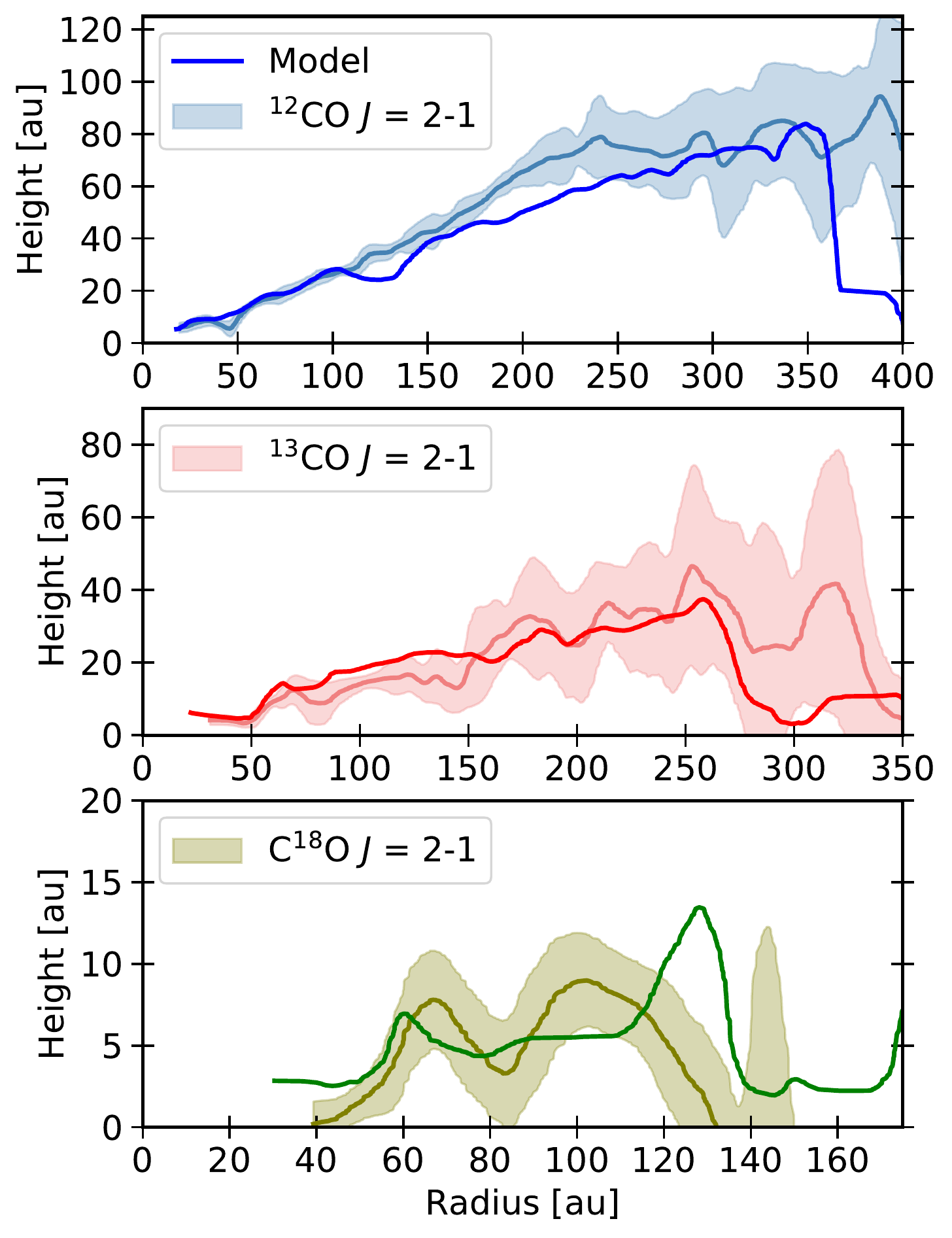}
    \caption{Derived emitting surfaces and calculated uncertainties of the 2-1 transitions of \CO{}, \thCO{} , and \CetO{}. Observed emitting surfaces with calculated error is shown in blue, red, and green from \citet{law20emit} while the model emitting surfaces are in the brighter blue, red, and green without accompanied error.}
    \label{fig:emit}
\end{figure}

An additional observational constraint on our model is the resolved emission heights of \CO{}, \thCO{}, and \CetO{} $J$ = 2-1 as measured in \citet{law20emit}. We calculate the emitting surfaces of \CO{}, \thCO{}, and \CetO{} using simulated image cubes and the same methods as applied to the observations by \citet{law20emit}. This method depends on the ability to resolve the front and back side of a given disk over multiple channels of an image cube, and assumes azimuthal symmetry and gas rotating in circular orbits \citep{Pinte18}. The emitting height extraction tools are found in the Python package \texttt{disksurf}\footnote{https://github.com/richteague/disksurf}. A comparison of the emitting layers of our model and observations are shown in Figure \ref{fig:emit}. 

\begin{figure*}
    \centering
    \includegraphics[scale=.5]{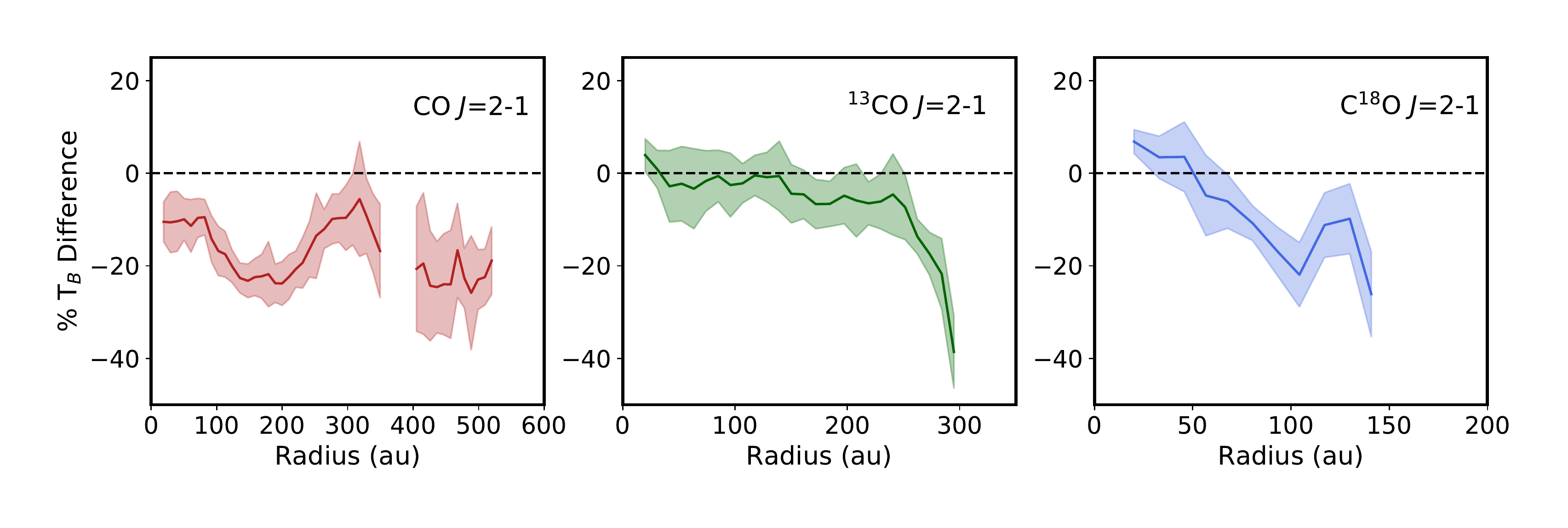}
    \caption{A comparison of the derived brightness temperature from the HD 163296 disk model and ALMA observations \citep{law20emit}. The y-axis shows the percentage difference between derived brightness temperature from the model and the ALMA results ($\frac{T_{\rm{rac2d}}-T_{\rm{obs}}}{T_{\rm{obs}}} \times 100$). The error is plotted as the shaded region and is motivated by the 1$\sigma$ uncertainty in the observed brightness temperature ($\frac{T_{\rm{rac2d}}-T_{\rm{obs}}\pm \Delta T_{obs}}{T_{\rm{obs}}} \times 100$). }
    \label{fig:emperical}
\end{figure*}

The calculated emitting layers for both simulated and observed data from \citet{law20emit} exist at similar heights, and agree within 1$\sigma$ at most radii. No additional changes to the model were necessary to arrive at this agreement between model and observation. 
Our modeled emission height for \CO{} $J$=2-1 reproduces what was derived using the ALMA observations up until $\sim$125 au; beyond this radius and up to 250 au, the model's \CO{} emits on average 11 au lower in the disk. Looking at \thCO{} emitting height, beyond $\sim$125 au the model reproduces the observed heights. This suggests that our model does not completely represent the CO vertical distribution, as it systematically produces a lower \CO{} $J$=2-1 emission height beyond $\sim$100au. However, this fit is quite good considering the detailed physics and chemistry included in the model. The \CetO{} emitting heights agree well with what is observed, especially within $\sim$100 au, where both our model and the observations pick up on some substructure, a bump at 70 au and subsequent dip at 85 au. Our model shows another increase in the \CetO{} emission height at 110 au, not present in the ALMA observations, a feature created due to relative CO abundance based on our depletion profile (see Figure \ref{COdep}).

\subsection{Comparison to Empirical Temperature Structure}\label{sec:emp}

In \citet{law20emit}, radial brightness temperatures (T$_{\rm{B}}$) are calculated for each of the CO $J$= 2-1 isotopologue lines, with a resolution of a quarter of the beam size. Here, we derive the brightness temperature of our model using an identical procedure to enable a consistent comparison. We compare the empirically derived temperatures for each of the $J$= 2-1 lines in Figure \ref{fig:emperical}. At most radii, the model's derived brightness temperature agrees within 10\% of the observed brightness temperature. Regions where the brightness model temperature is less than $\sim$20\% of the observed T$_{\rm{B}}$ corresponds to regions where the emitting heights diverge. For example, in the \CO{} comparison, the brightness temperature derived from the model is 20\% less than the observed value starting at $\sim$ 125 au, precisely where the \CO{} emitting heights diverge and the model \CO{} emitting height is 11 au deeper in the disk, where temperatures are cooler. Brightness temperatures measured from \thCO{} are in good agreement, up until $\sim$250 au. Beyond $\sim$250 au the model's emitting height sharply decreases while the observed emitting heights stay relatively constant (although with large uncertainty), which explains the large disconnect between the model and observed brightness temperatures. The average brightness temperature derived from ALMA observations of \CetO{} is 24 K, thus at radii where the model differs from the observed brightness temperature the most ($\sim$20\% less than observed), it is by just 4-5 K. Based on these comparisons, we show that this temperature structure based on a model derived by matching radial intensity profiles of CO, the disk SED, and unresolved fluxes of HD and upper transitions of CO, reproduces the observed brightness temperatures relatively well. 

\subsection{Implications for Disk Mass}\label{sec:diskmass}

This model uses a disk mass of 0.14 \Msun{}. However, there is a degeneracy between the CO and \Htwo{} surface densities. Increasing one parameter while decreasing the other by the same factor produces a model that reproduces nearly identical radial profiles in CO as the original model \citep{Calahan20}. Any change in heating based on the \Htwo{} or CO mass (heating via \Htwo{} formation, \Htwo{} and CO self-shielding) will not significantly affect the regions where the dust and gas temperatures are coupled. This applies to the vast majority of the disk, and where the bulk of the HD flux originates. Observing the flux of the molecule HD is a way to break this degeneracy and directly probe mass, as its ratio to \Htwo{} is well understood \citep{Linsky98}. Unfortunately, there was a non-detection of HD towards HD 163296 when observed with the PACS instrument on \textit{Herschel} \citep{Poglitsch10,Pilbratt10}. \citet{Kama20} provides 3$\sigma$ line flux upper limits for the HD 1-0 and 2-1 lines, $6.0 \times 10^{-18}$ W/m$^{2}$ and $3.0 \times 10^{-18}$ W/m$^{2}$ respectively. Our model predicts an HD 1-0 flux of $3.3 \times 10^{-18}$ W/m$^{2}$ and a 2-1 flux of 8.26 $\times 10^{-19}$ W/m$^{2}$. While it remains below the flux upper limit, this results in  considerable uncertainty in the mass. 

To determine an upper mass limit, we use our final model and increase the disk gas mass while decreasing the CO abundance by the same factor. Due to the degeneracy between \Htwo{} and CO, we can continue to reproduce the observed CO radial profiles while increasing the mass of the disk. An HD 163296 disk model with a mass of 0.35 \Msun{} produces a predicted HD 1-0 flux of 5.9 $\times 10^{-18}$ W/m$^{2}$ and 2-1 flux of 2.0 $\times 10^{-18}$ W/m$^{2}$. This is our upper mass limit, as constrained by the HD flux. This upper mass limit is higher than the majority of mass estimates for HD 163296, and lower than the largest estimate for HD 163296 which is 0.58 \Msun{} from \citet{Woitke19}. A more recent estimation for HD 163296 is presented in \citet{Booth19} using the optically thin CO isotopologue, $^{13}\rm{C}^{17}\rm{O}$, and predict a mass of  0.21 \Msun. This is a higher mass than we use in our model, but is also consistent with the HD flux limits and our thermo-chemical model. Using their derived HD flux limits, \citet{Kama20} determined an mass limit for the disk of $\leq$ 0.067 \Msun{}. They created a disk model that reproduced dust observations and a number of unresolved rotational transitions of \CO{} as well as a resolved observation of \CO{} $J$ = 3-2. There are two significant differences between our physical models that could explain the disparity between our calculated upper mass limits. Their HD 163296 model uses a star with a luminosity of 31 L$_{\odot}$ \citep{Folsom12} while this model uses a much lower luminosity of 17 L$_{\odot}$ \citep{Fairlamb15}. Additionally, they do not deplete CO which acts as a gas coolant, namely in the decoupled thermal regions, leading to a slightly cooler disk than ours. The contrast between our two models using the same HD flux observation to derive different disk mass estimates highlights the importance of a well-defined temperature structure.

\begin{figure}
    \centering
    \includegraphics[scale=.5]{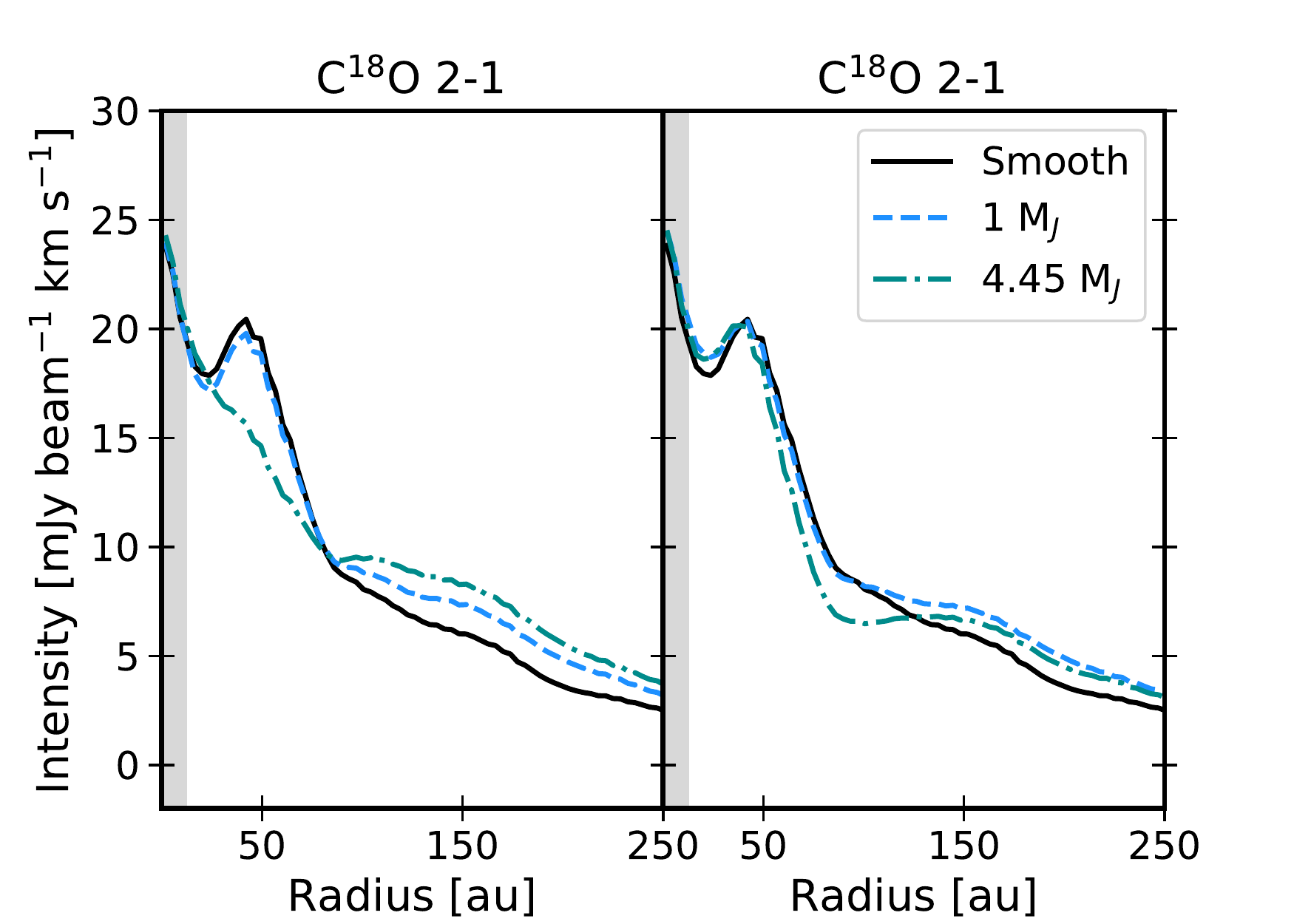}
    \caption{\CetO{} $J$ = 2-1 radial profiles for a smooth model (black), a model with a gap in the small dust and gas population that corresponds to a 1 M$_{J}$ planet (blue dashed), and 4.45 M$_{J}$ planet (teal dashed). The left plot shows the results for a gap centered at 48 au, and the right a gap at 86 au. }
    \label{fig:radprof_gap}
\end{figure}

\begin{figure}
    \centering
    \includegraphics[scale=.42]{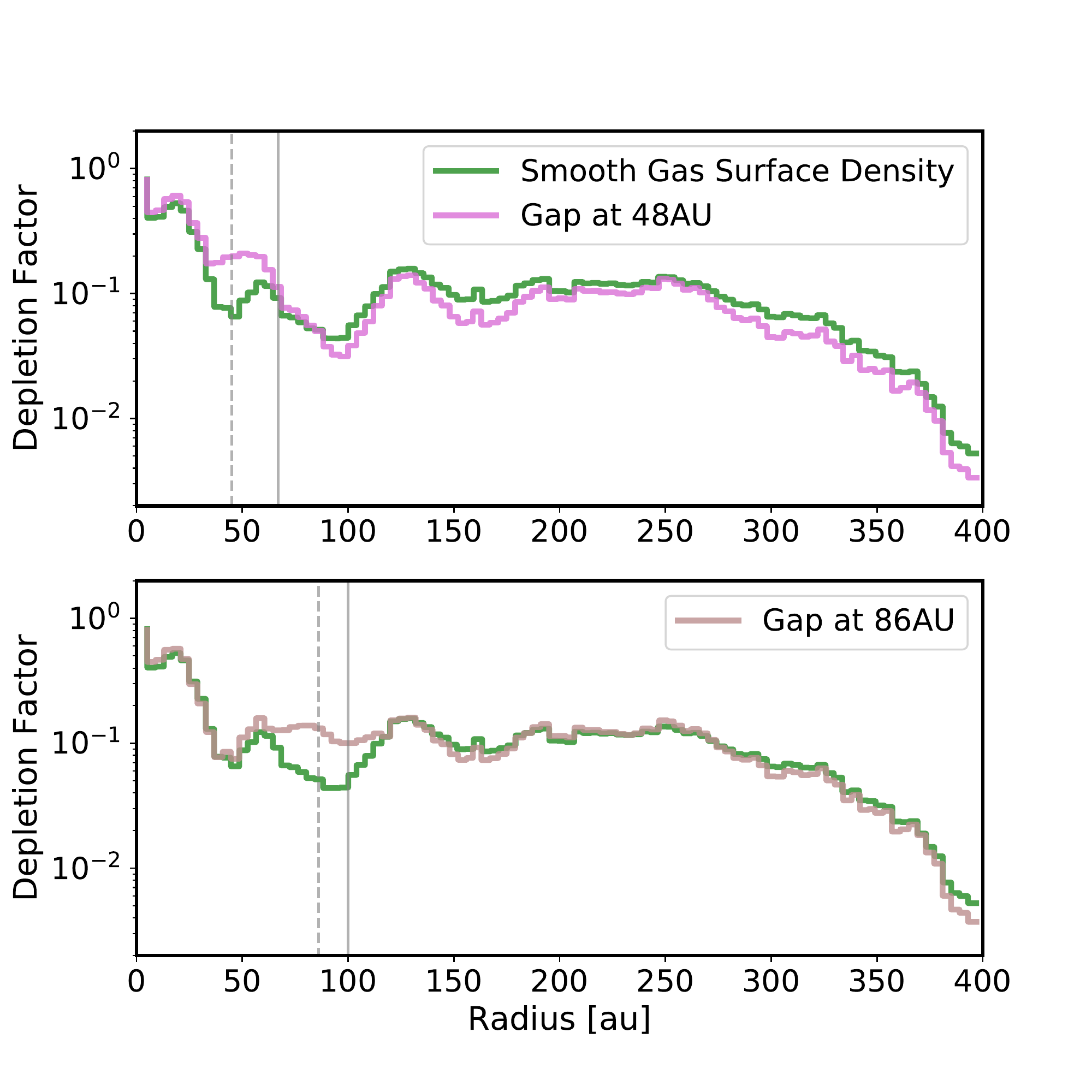}
    \caption{The calculated CO depletion for a smooth surface density model (green), and for a models with a gap at 48 au (top, pink) and 86 au (bottom, dark pink). The dashed grey lines correspond to the center of a gap, and the solid grey link corresponds to a ring. When imposing a gap in the total surface density, the CO depletion is flat or enhanced as opposed to being a local minimum at the location of the gap.}
    \label{fig:COdepgaps}
\end{figure}

\begin{figure}
    \centering
    \includegraphics[scale=.5]{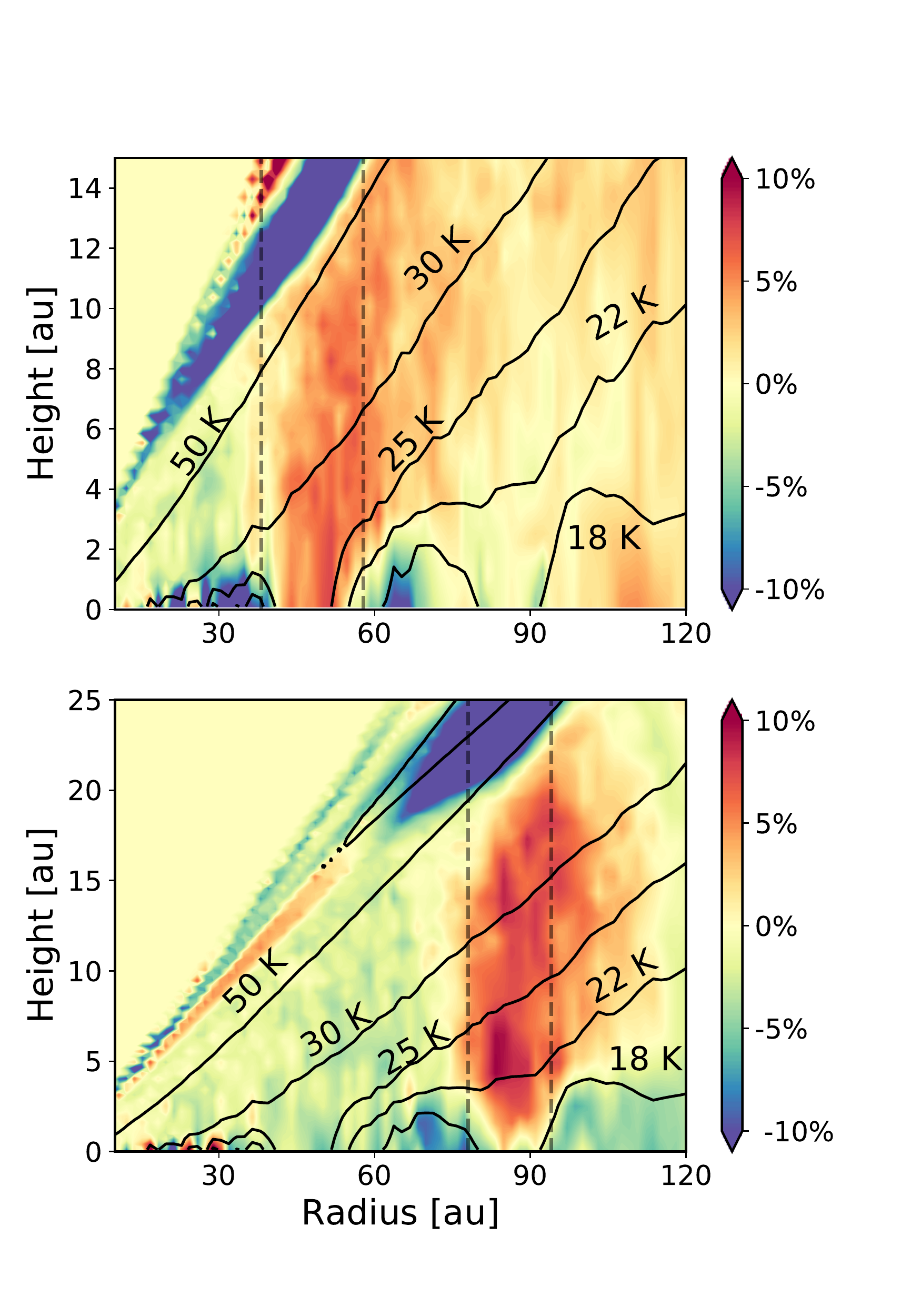}
    \caption{A comparison of gas temperatures between a model with a smooth gas surface density, and a model that has a gap in the gas at 48 au or 86 au. Both models have gaps in the large dust population and both models match observed radial profiles of CO, and follow a CO depletion profiles seen in Figure \ref{fig:COdepgaps}. Contour lines show gas temperatures of the smooth gas surface density model. The background corresponds to percentage difference from the smooth surface density model.  Positive values imply a hotter temperature in the gas gap models, negative values indicate hotter temperature in the smooth model. The dashed black lines correspond to the gap width.}
    \label{fig:gaptempcompare}
\end{figure}

\subsection{Implications for Gap Thermo-Chemistry}\label{sec:gaps}

The temperature of gas in potentially planet-carved gaps can provide insight into planet formation models directly linking protoplanetary disk environments to planet formation theories. The effects of disk geometry on temperature have been studied previously and it has been shown that the properties of local perturbations in the disk, including gap size and depth, radial location from the star, and disk inclination, will have an effect on the temperature structure \citep{Jang-Condell08}. Gaps have been found to be either cooler or warmer than the surrounding medium, dependent on disk geometry and other model assumptions. A decrease in the gas and small dust surface density exposes material in and near the midplane to more UV flux, allowing for an increase in the temperature of both the gas and dust \citep{vanderMarel18, Alarcon20}, however puffed up walls can produce shadows cooling the disk midplane \citep{Nealon19} or the gas-to-dust ratio can be low enough to decouple gas and dust temperatures heating only the dust \citep{Facchini18}. Using our final model, we explored the effect that corresponding gaps in the gas and small dust in HD 163296 may have on the presumed CO depletion, and subsequent gap temperature effects.

The gaps observed in the continuum emission in the HD 163296 disk are accounted for in the surface density of the large grains in our model. Meanwhile, the gas and small dust surface densities are smooth, and do not account for these gaps. This is a deliberate decision, as a primary goal of this study is to constrain the global thermal properties of the gas disk in general. %We find that the observed CO radial profiles have a local maximum at the gap locations (most notably at 48 au in \CetO{} $J$ = 2-1). 
We find that even with a smooth \Htwo{} gas distribution, the location of the gaps in the large dust population are warmer than outside of the gap by on order of a few Kelvin. This is supported by previous studies of the HD 163296 disk \citep{vanderMarel18,Rab20} which also find an increased temperature at gap locations using thermo-chemical models and observations of the gas and dust. %\citet{Rab20}, attributes the brightness peaks in CO to be not only from a warmer temperature in the gap, but more so from a combination of extra gas emission from the back side of the disk through the optically thin gap and absorption of line emission from the back side of the disk through dust rings accompanying the gaps. 

Gaps are often attributed to on-going planet formation. To determine the gas and small dust depletion level within the gap, we created four models representing two of the gaps and two possible planet masses. The most prevalent gaps in the continuum (the widest, and highest contrast gaps) are located at 48 au and 86 au. The location, depth, and width of the gas and small dust gap used in our model is motivated by the measured gap widths in continuum from \citet{Isella18} and the predicted planet masses from \citet{Zhang18} dependent on dust size distribution and viscosity. We set gap depths, parameterized by a depletion of \Htwo{} gas within the gap, using two assumed planet masses, 1 M$_{J}$, which represents a typical planet mass estimate for the HD 163296 disk \citep{Pinte18_hd16, Teague20}, and 4.45 M$_{J}$, which represents the highest mass planet predicted within HD 163296 \citep{Zhang18}. We used Equation 5 in \citet{Kanagawa15} to determine gap depth using our model's density distribution, viscosity ($\alpha$ = 10$^{-2}$), and planet mass. A 1 M$_J$ planet at 48 au leads to a gas depletion depth of 23\%, and at 86 au, 19\%. A 4.45 M$_J$ planet at 48 au leads to a gas depletion depth of 85\%, and at 86 au, 82\%. 

We use our final HD 163296 disk model and re-run it with the new gas surface density dependent on gap location and planet mass. The two models with a gap from a 1 M$_{J}$ planet produces a negligible change in the observed radial intensity profile while a model with gap depths corresponding to 4.45 M$_{J}$ planet shows a significant decrease in flux, see Figure \ref{fig:radprof_gap}. We then calculate a new CO depletion profile in order for the models with a 4.45 M$_{J}$ planet induced gap to match the observed radial emission profile. Those profiles are show in Figure \ref{fig:COdepgaps}. Previously, with a smooth gas distribution, the CO depletion profile had local minima at the location of the gaps. In the case of a gas and small dust surface density with deep gaps, the CO abundance increases at the location of the gaps bringing it to a level on par with the depletion factors just outside of the gaps. Our method of CO depletion makes it difficult to disentangle chemical/gap effects on the CO abundance, so it is impossible to decern if this relative increase in CO in the gaps is an enhancement or a leveling off to a more constant CO depletion across radius. Studies such as \citet{Alarcon20} and \citet{vanderMarel16} have predicted that CO enhancement in the gaps is needed to match observations. One possible mechanism to enhance the CO abundance is the presence of meridional flows found at the gaps in HD 163296 by \citet{Teague19}, which can transport gas and small grains from upper atmosphere, bringing CO sublimating from grains and a rich chemistry to an otherwise chemically inert midplane. Whether or not CO is enhanced locally, is dependent on the depth of the gas gap. At present, the best method to determine gas depletion in gaps is to use kinematics to constrain the \Htwo{} pressure gradients \citep{Teague18} and thus the amount of CO chemical processing that might be happening \citep{alarcon21}

Using the new CO depletion profiles for our deep gas gap models, we find that the gap temperature increases by upwards of 10\% compared to a model with a smooth gas surface density. Figure \ref{fig:gaptempcompare} shows the difference in temperature between our smooth gas model, and our gapped models, both of which match the observed CO radial profiles. The increase in temperature in the gap arises due to the decreased UV opacity from the depletion of small grains, allowing more flux to enter the region. There is a region in the atmosphere above the heated layer that is cooler than a smooth gas surface density model by over 10\%. This occurs at the undershoot region where the gas temperature is lower than the dust temperature due to atomic lines becoming major coolants. In the gas gap models, this cooler, optically thick layer is at a slightly lower height.  

\begin{figure}
    \centering
    \includegraphics[scale=.55]{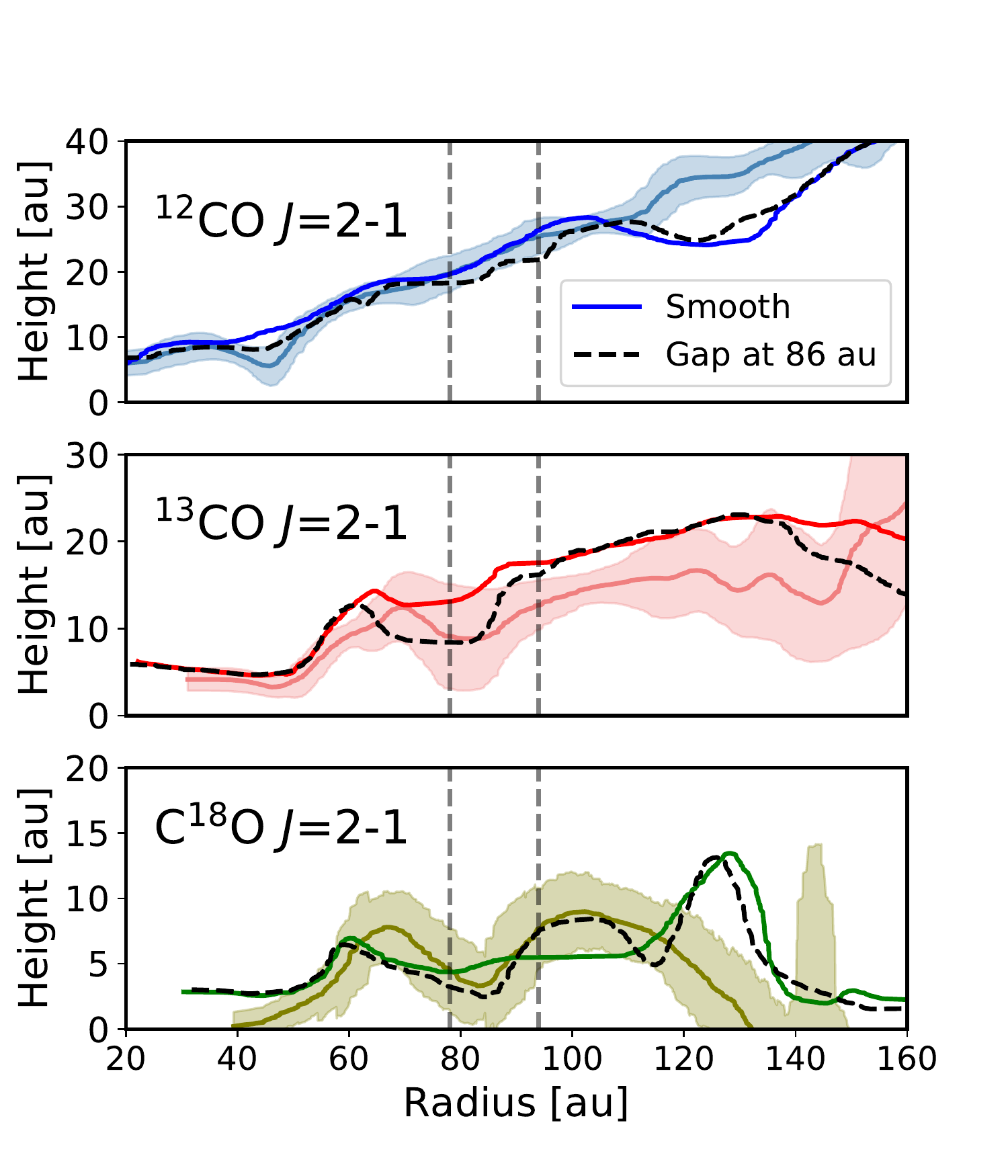}
    \caption{Emitting surface of \CO{}, \thCO{}, and \CetO{}  $J$=2-1 in a model that uses a depletion of CO (solid line), and a model that has a gap in the gas surface density at 86 au (dashed black line) as compared to observations (thick line).}
    \label{fig:c18oemit}
\end{figure}

We also compare the observed emitting surface of the CO $J$ = 2-1 line to the model with a deep gap at 86 au in Figure \ref{fig:c18oemit}. There is limited emitting surface information for \CetO{} $J$ = 2-1 at the location of the 48 au gap, thus we focus on the gap farthest out. The models with 1 M$_{J}$ planet gaps showed very little variation compared to the smooth model. The 4.45 M$_{J}$ model provides the best insight into how a significant deviation in the gas surface density could affect the emitting surface and the degeneracy between the \Htwo{} surface density and CO abundance. At the location of the gap, each CO isotopologue 2-1 transition emits from a lower layer, with \thCO{} and \CetO{} showing the strongest change in height ($<$ 6 au). This suggests that in highly gas-depleted gaps the degeneracy between the \Htwo{} surface density and CO abundance can be broken. An even more significant difference between the two models is highlighted in the \CetO{} $J$ = 2-1 emitting surface. Just beyond the gap at 86 au, the model with a deep gap in the surface density follows the observed emitting surface more so than the model with a smooth surface density and structured CO depletion profile. Exploring the limits of the degeneracy between CO and \Htwo{} is beyond the scope of this study, however extracting and comparing observed and simulated emitting surfaces may be an interesting tool to use in the future.

\section{Conclusion}\label{sec:conclusion}

Using high spatial resolution observations of \CO{}  $J$=2-1, \thCO{}  $J$=2-1, 1-0, \CetO{}  $J$=2-1, 1-0, and \CsvO{} 1-0 from the ALMA MAPS large program in concert with the thermo-chemical code RAC2D, we derive a 2D thermal structure for the disk around Herbig Ae star HD 163296. Our conclusions are the following:

\begin{enumerate}
    \item We derived a 2D thermal structure for the disk around Herbig Ae star HD 163296 that reproduces six spatially resolved rotational transition lines of CO and its isotopologues, in addition to the observed SED, structures observed in the continuum, the predicted HD flux remains below the observed upper limit, and we reproduce observed fluxes of 19 higher J level transitions of \CO{}.
    
    \item The derived temperature agrees well with empirically derived temperatures and calculated emitting heights. This temperature structure represents the full 2D thermal structure, filling in temperature information at spatial locations which observations do not directly probe.
    
    \item We calculate a CO depletion profile which shows a relative enhancement of CO within the CO snowline, a slightly lower depletion value between the CO snowline and 250 au, and then a significant drop of that can be explained via CO chemical processing.
    
    \item Using the derived thermal structure, we predict a midplane snowline location for CO of 60 au which corresponds to a freeze out temperature of 18 K. We also find an upper mass limit of 0.35 \Msun. 
    
    \item The temperature is locally enhanced in the mm continuum gaps at 48 and 86 au. This is true both for gaps consisting of only large grain depletion, and those with large, small grain and gas depletion. The temperature within the gap increases slightly when gas and small dust are depleted in addition to large grains, by at most 10\% in the case of significant depletion.   
\end{enumerate}

%\acknowledgments
\section{Acknowledgments}
This paper makes use of the following ALMA data: ADS/JAO.ALMA\#2018.1.01055.L. ALMA is a partnership of ESO (representing its member states), NSF (USA) and NINS (Japan), together with NRC (Canada), MOST and ASIAA (Taiwan), and KASI (Republic of Korea), in cooperation with the Republic of Chile. The Joint ALMA Observatory is operated by ESO, AUI/NRAO and NAOJ. The National Radio Astronomy Observatory is a facility of the National Science Foundation operated under cooperative agreement by Associated Universities, Inc.

J.K.C. acknowledges support from the National Science Foundation Graduate Research Fellowship under Grant No. DGE 1256260 and the National Aeronautics and Space Administration FINESST grant, under Grant no. 80NSSC19K1534. 
E.A.B. acknowledges support from NSF AAG Grant \#1907653.
K.Z. acknowledges the support of the Office of the Vice Chancellor for Research and Graduate Education at the University of Wisconsin – Madison with funding from the Wisconsin Alumni Research Foundation, and support of the support of NASA through Hubble Fellowship grant HST-HF2-51401.001. awarded by the Space Telescope Science Institute, which is operated by the Association of Universities for Research in Astronomy, Inc., for NASA, under contract NAS5-26555. 
K.R.S. acknowledges the support of NASA through Hubble Fellowship Program grantHST-HF2-51419.001, awarded by the Space Telescope Science Institute,which is operated by the Association of Universities forResearch in Astronomy, Inc., for NASA, under contract NAS5-26555.
K.I.\"O. acknowledges support from the Simons Foundation (SCOL \#321183) and an NSF AAG Grant (\#1907653). 
V.V.G. acknowledges support from FONDECYT Iniciaci\'on 11180904 and ANID project Basal AFB-170002.
F.A. acknowledges support from NSF AAG Grant \#1907653.
R.T. and F.L. acknowledges support from the Smithsonian Institution as a Submillimeter Array (SMA) Fellow.
J.B.B. acknowledges support from NASA through the NASA Hubble Fellowship grant \#HST-HF2-51429.001-A, awarded by the Space Telescope Science Institute, which is operated by the Association of Universities for Research in Astronomy, Inc., for NASA, under contract NAS5-26555.
A.S.B acknowledges the studentship funded by the Science and Technology Facilities Council of the United Kingdom (STFC).
J.D.I. acknowledges support from the Science and Technology Facilities Council of the United Kingdom (STFC) under ST/T000287/1.
C.J.L. acknowledges funding from the National Science Foundation Graduate Research Fellowship under Grant DGE1745303.
R.L.G. acknowledges support from a CNES fellowship grant.
M.L.R.H. acknowledges support from the Michigan Society of Fellows
Y.A. acknowledges support by NAOJ ALMA Scientific Research Grant Numbers 2019-13B and Grant-in-Aid for Scientific Research (S) 18H05222.
S. M. A. and J. H. acknowledge funding support from the National Aeronautics and Space Administration under Grant No. 17-XRP17 2-0012 issued through the Exoplanets Research Program. Support for this work was provided by NASA through the NASA Hubble Fellowship grant \#HST-HF2-51460.001-A awarded by the Space Telescope Science Institute, which is operated by the Association of Universities for Research in Astronomy, Inc., for NASA, under contract NAS5-26555.
A.D.B. acknowledges support from NSF AAG Grant \#1907653.
J.B. acknowledges support by NASA through the NASA Hubble Fellowship grant \#HST-HF2-51427.001-A awarded  by  the  Space  Telescope  Science  Institute,  which  is  operated  by  the  Association  of  Universities  for  Research  in  Astronomy, Incorporated, under NASA contract NAS5-26555.
C.W.~acknowledges financial support from the University of Leeds and from the Science and Technology Facilities Council (grant numbers ST/R000549/1 and ST/T000287/1).
H.N. acknowledges support by NAOJ ALMA Scientific Research Grant Code 2018-10B and Grant-in-Aid for Scientific Research 18H05441.
I.C. was supported by NASA through the NASA Hubble Fellowship grant HST-HF2-51405.001-A awarded by the Space Telescope Science Institute, which is operated by the Association of Universities for Research in Astronomy, Inc., for NASA, under contract NAS5-26555. 
F.M. acknowledges support from ANR of France under contract ANR-16-CE31-0013 (Planet Forming Disks) and ANR-15-IDEX-02 (through CDP ``Origins of Life").
G.C. is supported by the NAOJ ALMA Scientific Research Grant Number 2019-13B.
Y.Y. is supported by IGPEES, WINGS Program, the University of Tokyo.

We also thank the referee for a careful read of the text and insightful comments.

\facilities{ALMA}
\software{RAC2D \citep{Du14}, bettermoments \citep{Teague18}, GoFish \citep{Teague19}, disksurf (https://github.com/richteague/disksurf) , CASA \citep{McMullin07}, RADMC3D \citep{radmc3d}},  scipy \citep{scipy}, Matplotlib \citep{Matplotlib}, Astropy \citep{astropy:2013,astropy:2018}

\appendix
\restartappendixnumbering

\section{Parameter Exploration Analysis}\label{sec:param}

\begin{figure}
    \centering
    \includegraphics[scale=.45]{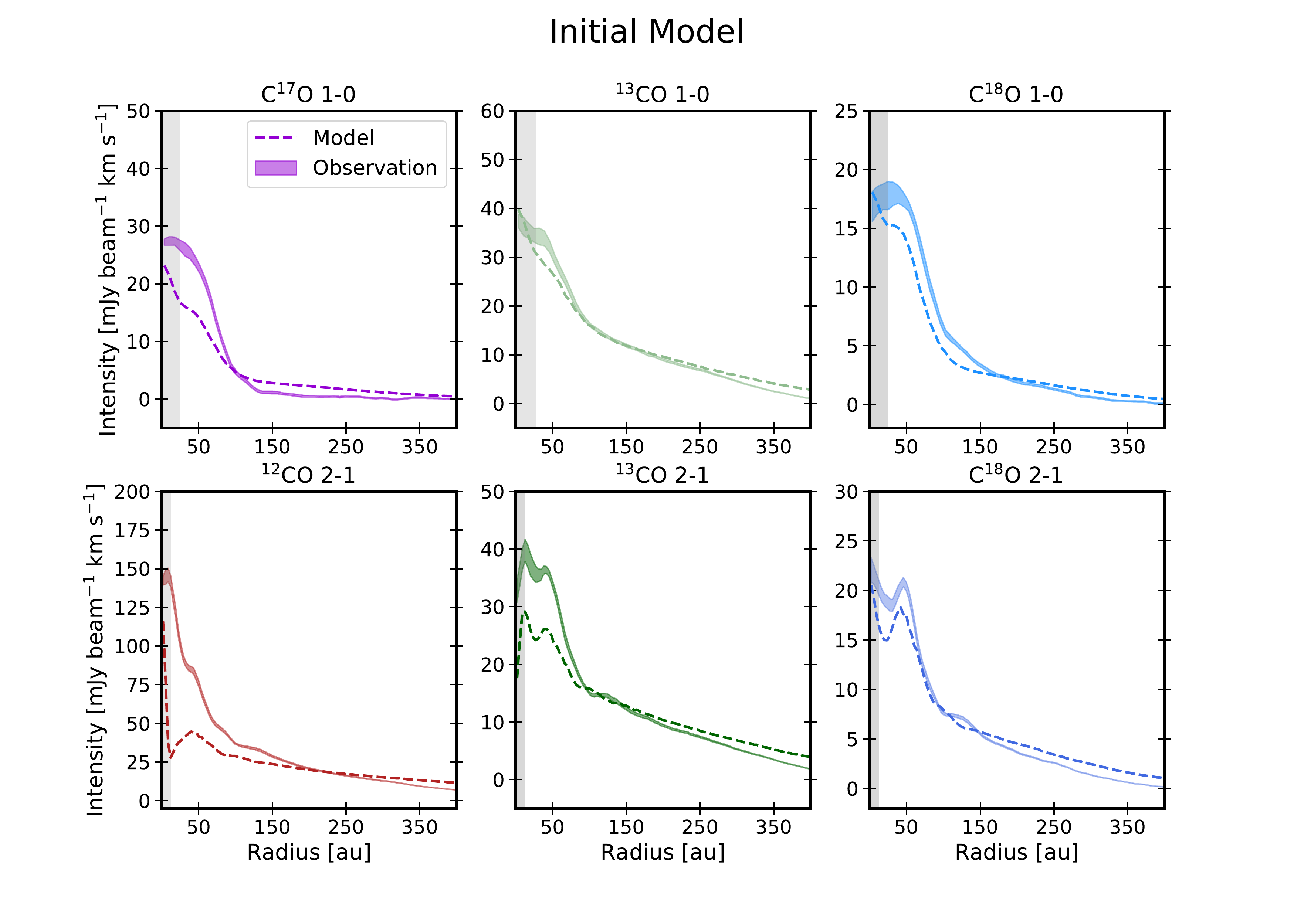}
    \caption{Radial emission profiles as predicted by the initial HD 163296 model based on Z21 (Z21) compared to the observed profiles (solid line). Parameters are listed in Table \ref{params}. This model uses the CO depletion calculated in Z21, applied before the temperature calculation, the chemistry runs for 0.01 Myr, and this model lacks the excess heating found necessary to reproduce the \CO{} $J$=2-1 observation}
    \label{fig:init}
\end{figure}

Our model of the HD 163296 disk based on the density structure from Z21 with an updated CO depletion profile reproduces radial profiles from all lines except \CO{} $J$=2-1 relatively well (see Figure \ref{fig:init}). Thus, this initial model appears to roughly represent the 2D thermal structure, but cannot reproduce the disk layers at which \CO{} $J$=2-1 emission originates, as this line is underpredicted by a factor of about two within the brightest region (within 50 au). However, improvements can be made to all lines except for \CetO{} $J$=2-1 (and arguably \thCO{} 1-0) because they are slightly underpredicted in the inner 75 au, while slightly overpredicted in the outer disk. In order to achieve a more complete and accurate thermal profile, it is worth exploring the sensitivity of the radial emission profiles to the disk physical parameters. 

Throughout the parameter exploration, we did not alter the CO depletion profile. We find that the CO depletion is only degenerate with disk physical parameters that affect the total gas mass or mass distribution ($\gamma$). When the scale height, characteristic radius, or flaring parameter are changed, we find that there is no single CO depletion profile that brings all lines of CO towards what is observed.

We ran a set of models that explored the parameter space of gas mass, small dust mass, flaring parameter ($\Psi$), surface density power-index ($\gamma$), characteristic radius (r$_{\rm c}$), critical height (h$_{\rm c}$), and radial extent and allowed for the temperature to evolve in a new physical environment. Because HD 163296 has been widely studied, we determined the range exploration based on literature values. There has been a wide array of gas mass estimates, from as low as $8 \times 10^{-3}$  up to 0.58 \Msun{} \citep{Isella07, Williams14, Boneberg16, Miotello16, Williams16, Woitke19, Booth19, Powell19, Kama20}. Our initial gas mass is 0.14 \Msun{}, which is on the higher end, thus we created three models that are lower in mass, and one higher in mass: $8 \times 10^{-3}$, $1 \times 10^{-2}$, $6.7 \times 10^{-2}$ (predicted gas upper limit based on the HD 1-0 flux \citep{Kama20}), and 0.21 \Msun{} (the HD 163296 disk mass estimate from \citet{Booth19}).  In terms of small dust mass, the SED constrains the range of exploration. We start with a mass of $1 \times 10^{-4}$ \Msun, and explore values above and below this with the extremes being clear under and over predictions of the SED flux beyond 10$\mu$m $\micron$: $1\times 10^{-6}$ \Msun, $1\times 10^{-5}$ \Msun, $1\times 10^{-3}$ \Msun, $1\times 10^{-2}$ \Msun. Flaring for HD 163296 has also had a wide range of estimates throughout the literature, and different studies assume both flat and flared disks. We explore models that use $\Psi$ values above and below what we have used (1.08). This includes 0.05, 1.0, 1.6, 2.2 \citep{Tilling12, deGregorio13, Woitke19, Kama20}. The surface density index, $\gamma$, has a natural limit in our description of surface density (see Equation 1). We explore values above and below our initial value of 0.8: 0.2, 0.6, 1.2, and 1.8. There is \CO{} $J$=2-1 detected out to 600 au, thus we only explored values above that for r$_{\rm out}$: 700, 1000, and 1200 au. For r$_{\rm c}$,  h$_{\rm c}$ we explore four values, the lowest of which corresponds to the mm-dust distribution, and the largest being double our initially inferred value.

\begin{figure}
\centering
 \includegraphics[scale=.45]{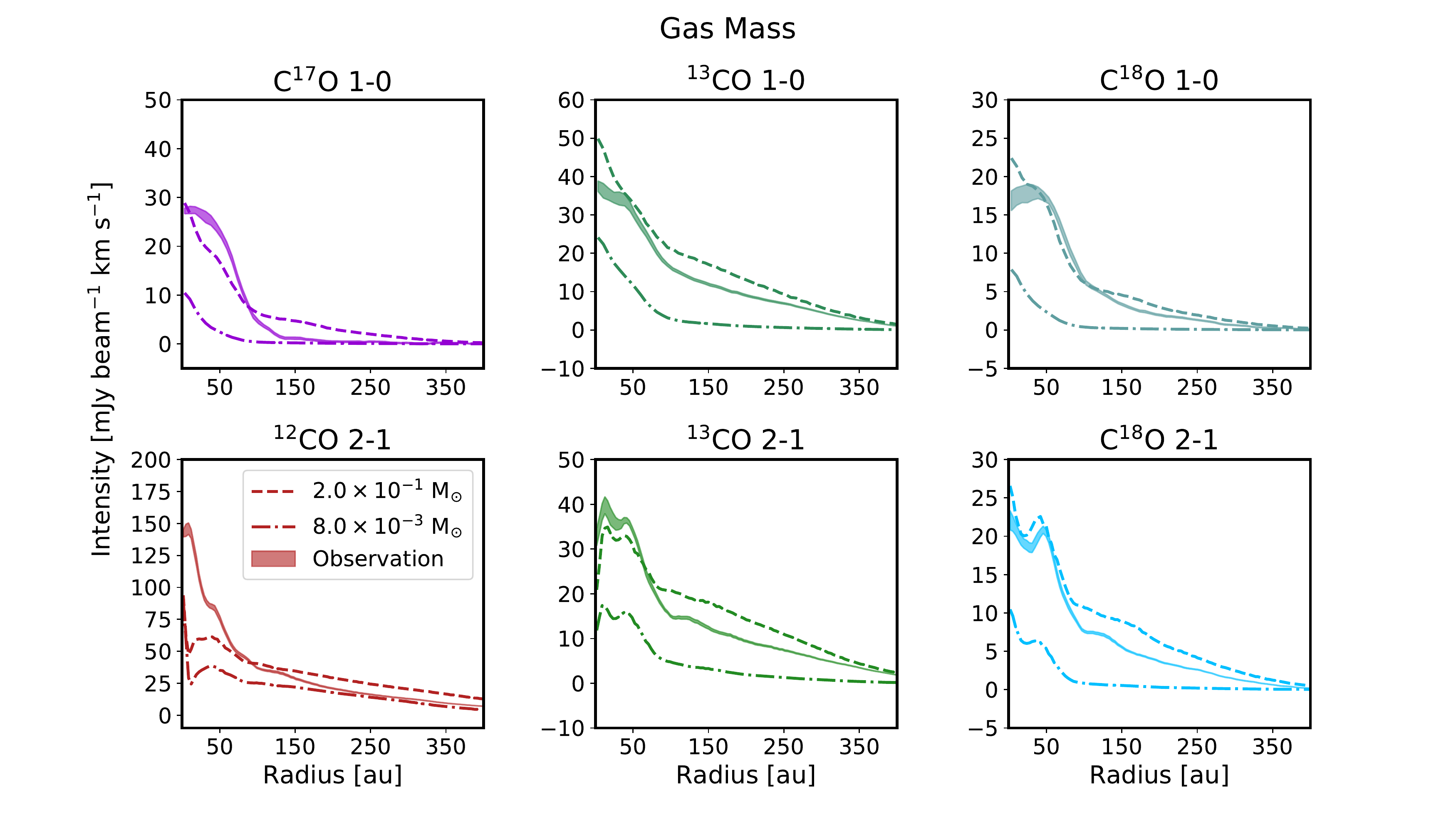}
\caption{Modeled radial emission profiles of HD 163296 model based on Z21 compared to the observed profiles (solid line). These models exhibit varying gas mass: 0.2 \Msun{} which is a mass prediction for HD 163296 by \citet{Booth19} and 0.008 \Msun{} which is the smallest predicted mass in literature.}
\label{fig:gasmass}
\end{figure}

\begin{figure}
    \centering
    \includegraphics[scale=.45]{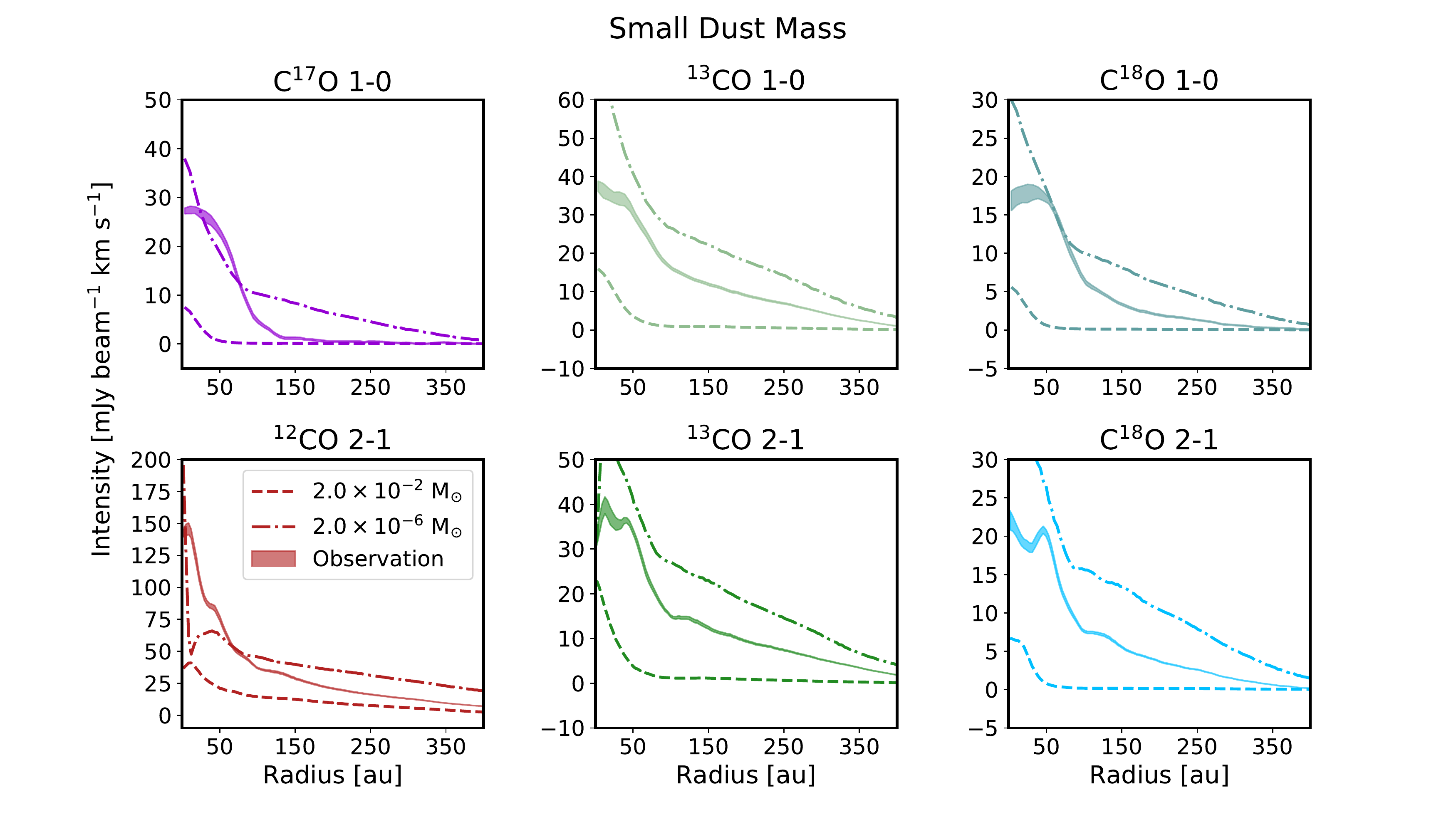}
    \caption{Modeled radial emission profiles of HD 163296 model based on Z21 compared to the observed profiles (solid line). These models vary in small dust mass: $2 \times 10^{-6}$ \Msun{} and $2 \times 10^{-2}$ \Msun{}. The small dust mass is constrained by the SED, and these limits represent dust mass values that are just under and over predicting the SED.}
\label{fig:smalldust}
\end{figure}

\begin{figure}
    \centering
    \includegraphics[scale=.45]{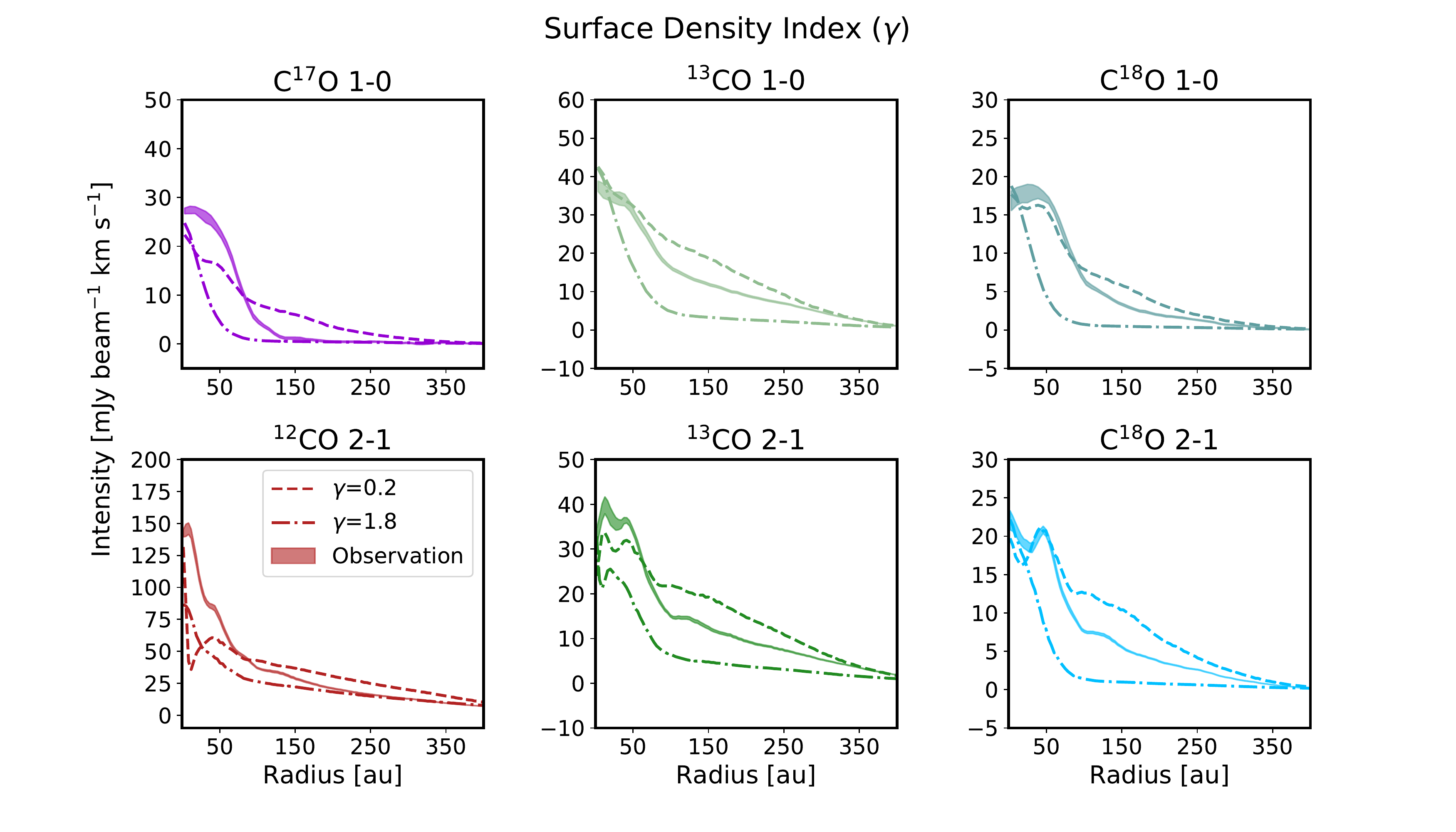}
    \caption{Modeled radial emission profiles of HD 163296 model based on Z21 compared to the observed profiles (solid line). The models vary in surface density index, $\gamma=$0.2 and 1.8. There is a limit for $\gamma$ of 0 - 2 based on our definition of surface density, thus we explored values at the upper and lower ends of that natural range.}
\label{fig:gamma}
\end{figure}

\begin{figure}
    \centering
    \includegraphics[scale=.45]{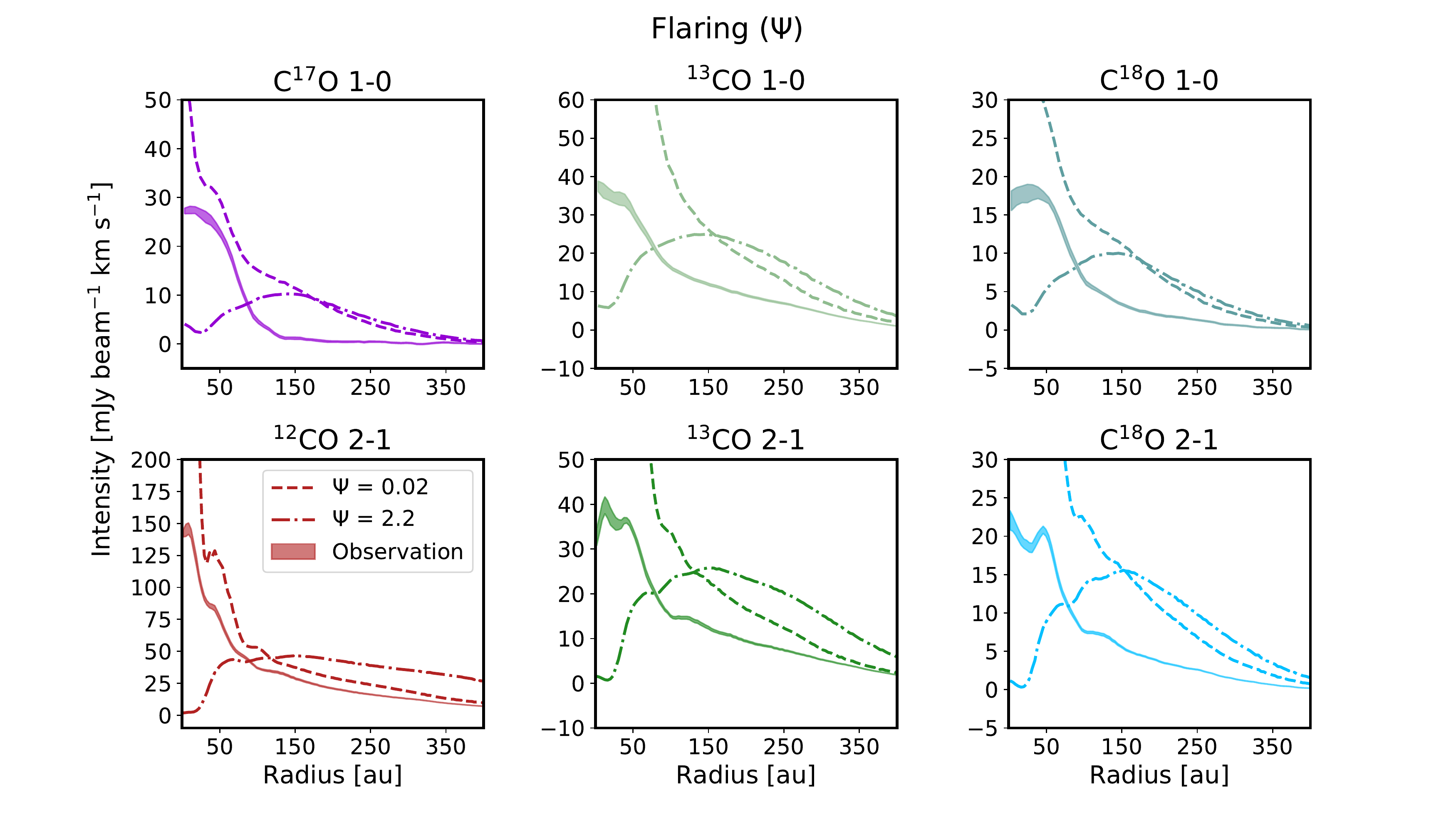}
    \caption{Modeled radial emission profiles of HD 163296 model based on Z21 compared to the observed profiles (solid line). These models exhibit varying gas mass: 0.2 \Msun{} which is a mass prediction for HD 163296 by \citet{Booth19} and 0.008 \Msun{} which is the smallest predicted mass in literature.}
 \label{fig:psi}
\end{figure}

\begin{figure}
    \centering
    \includegraphics[scale=.45]{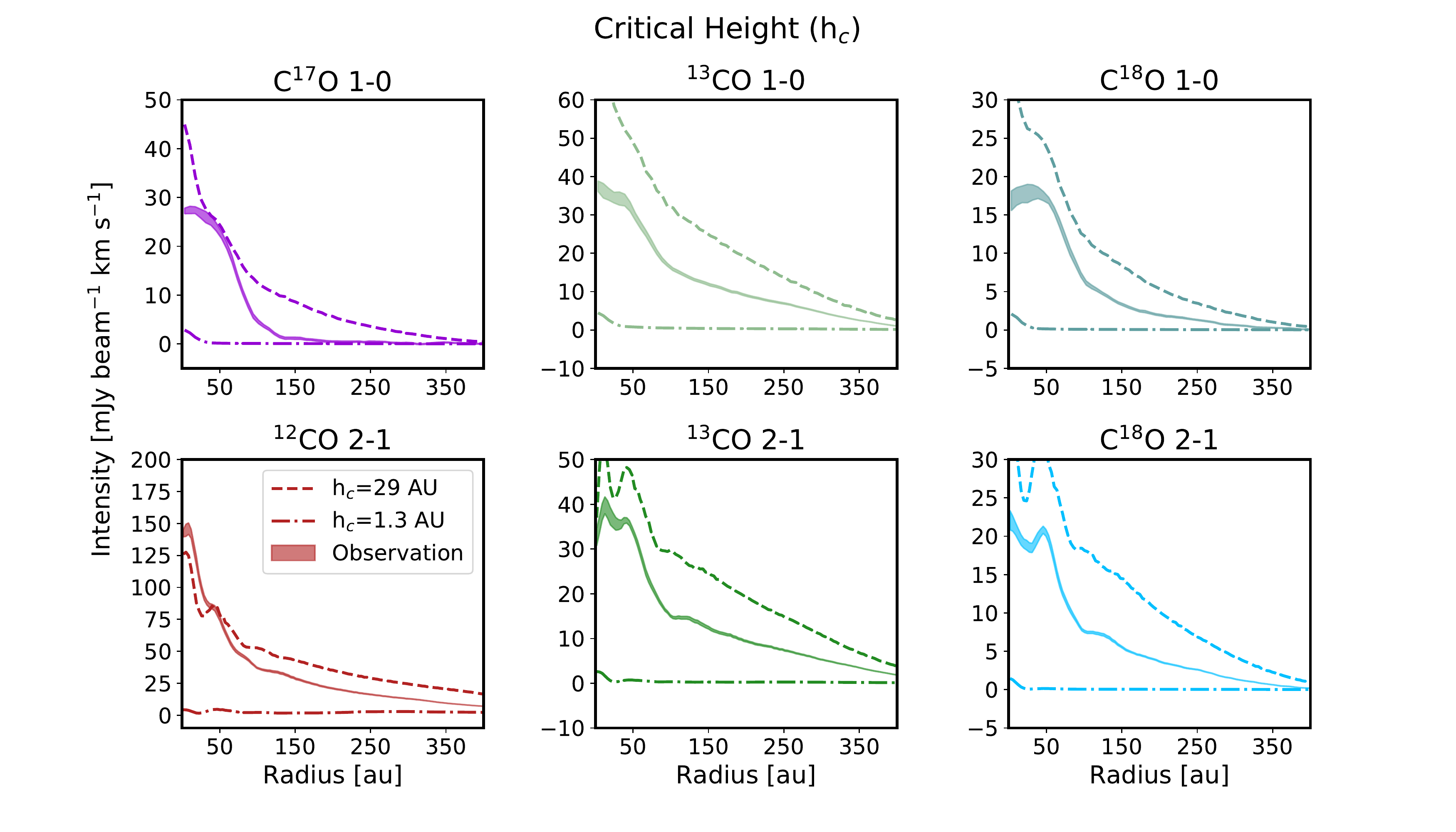}
    \caption{Modeled radial emission profiles of HD 163296 model based on Z21 compared to the observed profiles (solid line). These models vary in scale height. We explore scale height values as low as what has been used to describe the large grain population, to as high as twice as what we originally predicted.}
\label{fig:height}
\end{figure}

\begin{figure}
    \centering
    \includegraphics[scale=.45]{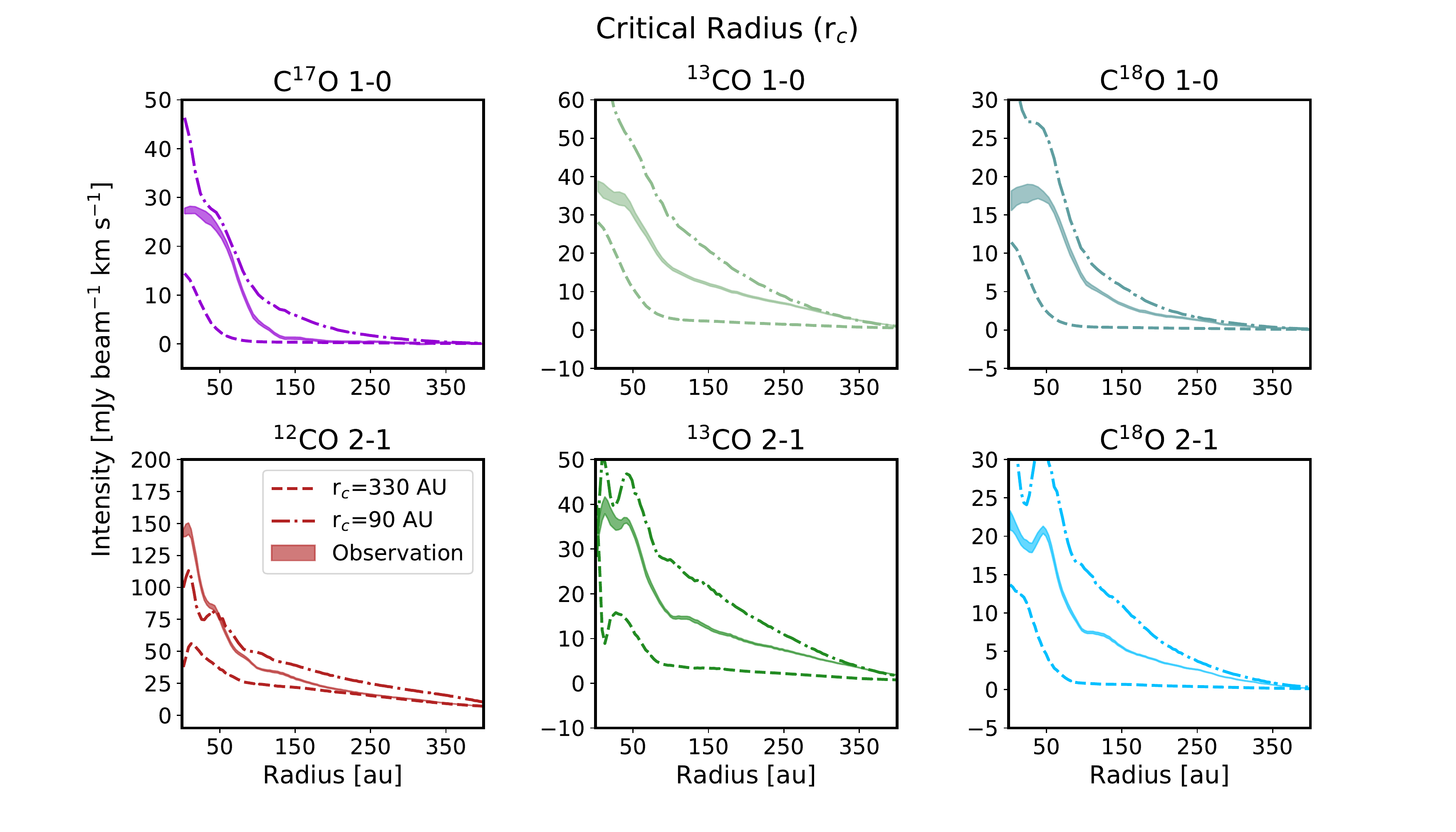}
    \caption{Modeled radial emission profiles of HD 163296 model based on Z21 compared to the observed profiles (solid line). These models vary in characteristic radius. We explore characteristic radius values as low as what has been used to describe the large grain population, to as high as twice as what we originally predicted.}
\label{fig:radius}
\end{figure}

\begin{figure}
    \centering
    \includegraphics[scale=.45]{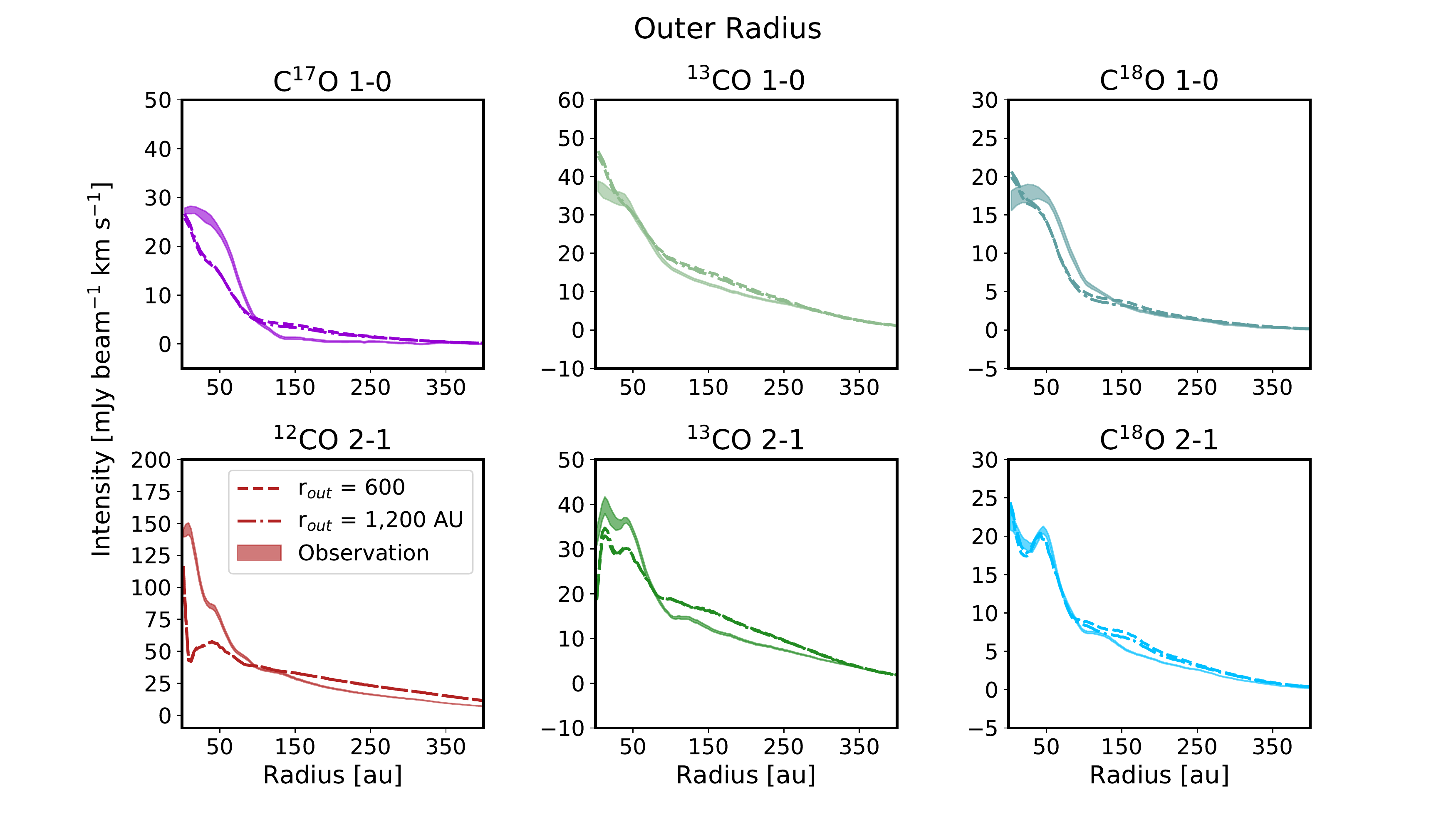}
    \caption{Modeled radial emission profiles of HD 163296 model based on Z21 compared to the observed profiles (solid line). These models vary in outer radius. We explore radius values as low as 600 au, which is the largest radial extent of the observed \CO{}, to twice that size.}
\label{fig:rout}
\end{figure}

Changes in gas mass, small dust mass and scale height have similar effects on the CO radial profiles, increasing or decreasing the CO intensity with increasing/decreasing gas mass and scale height and decreasing/increasing small dust mass. Changing the scale height will increase or decrease the CO flux along all radii for all transitions (see \ref{fig:height}), while small dust mass and gas mass effect some lines more than others (see Figures \ref{fig:gasmass} and Figures \ref{fig:smalldust}). For example, changes in the mass of the gas or small dust populations do not have as strong an effect on \CO{} $J$ 2-1 as it does on nearly every other line. While mass changes in these two populations appear to have similar effects, any degeneracy between the two can be broken as the small dust population is constrained by the SED. As the flaring parameter, $\Psi$, increases,  emission is enhanced in the outer disk (the divide between `outer' and `inner' disk depends on the characteristic radius). As $\Psi$ decreases, there is a significant increase in the inner disk (see Figure \ref{fig:psi}). We found for this model that even small changes in $\Psi$ strongly affect the final radial profile across all lines. The surface density power-index, $\gamma$, tends to leave the flux in the inner few au unaffected, but with a smaller $\gamma$ more emission can be found farther out in the disk (see Figure \ref{fig:gamma}). This is due to the fact that a smaller $\gamma$ produces a population that is more evenly distributed. The characteristic radius affects both the height and surface density of a given population, with lower r$_{\rm c}$ values producing a rapid increase in height and a turn over the surface density at a shorter radius (see Figure \ref{fig:radius}). The combination of these two effects produces radial profiles that are brighter for smaller critical radii. The outer radius values we explored did not produce significantly different CO radial profiles, due to the fact that the majority of the emission from all lines exists within 400 au, thus an outer radius cut off well beyond this limit does not affect the observed emission (see Figure \ref{fig:rout}).

After this exploration of parameter space, and subsequently creating a number of models that altered multiple parameters simultaneously, we determined that the best set of parameters were the ones that were used by Z21. Keeping with these values, the derived thermal structure matches five out of six of the radial intensity profiles relatively well and is the best $\chi^{2}$ fit for the SED (per Z21).

\section{Gas Temperature Structures in HD 163296 Models from the Literature}\label{sec:compare}

\begin{figure*}
    \centering
    \includegraphics[scale=.5]{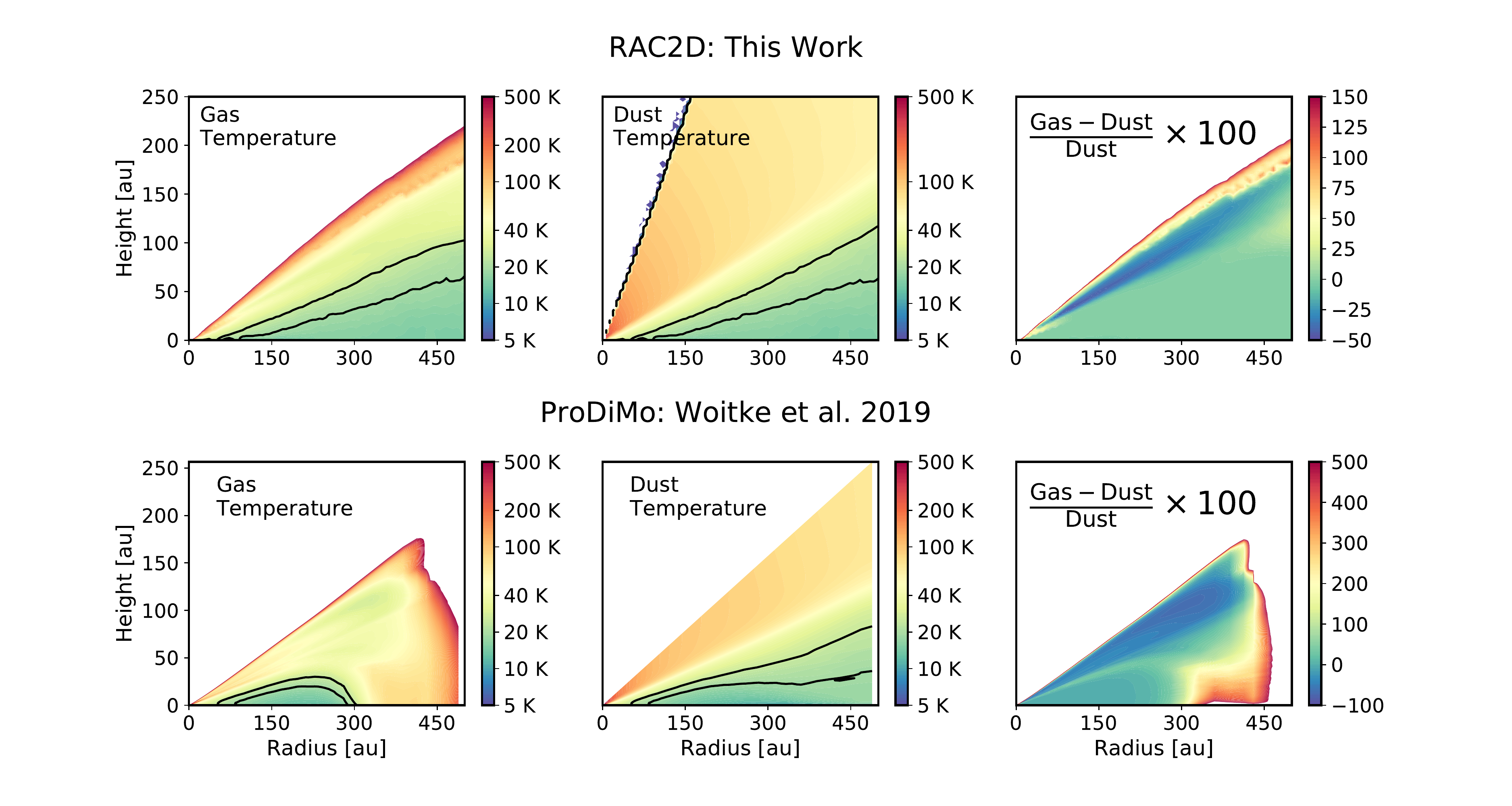}
    \caption{A comparison between this work's HD 163296 thermal structure that best reproduces the CO radial profiles, and the HD 163296 specific model from \citet{Woitke19}. The two contours in black follow 19K and 25K.}
    \label{fig:compare}
\end{figure*}

There has been one other recent attempt to characterize the 2D thermal structure of the HD 163296 disk specifically, using a thermo-chemical code, matching multiple line observations and the SED. Those results are presented in \citet{Woitke19}. They use the thermo-chemical code ProDiMo \citep[PROtoplanetary DIsk MOdel][]{Woitke09,Kamp10,Woitke16} and derive a disk model that reproduces observed line fluxes from infrared to millimeter wavelengths within a factor of about two, along with the observed SED. The model outputs from \citet{Woitke19}
are available publicly, and we compare our final thermal structure to their results in Figure \ref{fig:compare}. The dust temperatures in both models are very similar, with the disk in our model being slightly more flared. The gas temperatures are very similar within $\sim$ 200 au, while past 300 au the ProDiMo model has a much hotter disk than what this study predicts. That relatively hot temperature most likely would affect the spatially resolved radial profiles, namely \CO{} $J$=2-1 which emits beyond 300 au. 

There are a few possible reasons as to why that study did not find it necessary to invoke additional heating in the layers at which \CO{} $J$=2-1 primarily emits. A key difference between the two models is the underlying gas mass; 0.58 \Msun{} from \citet{Woitke19} vs. 0.14 \Msun{} in this study (although the ProDiMo model used a pre-Gaia distance of 119 pc, as opposed to the Gaia determined 101 pc). Additionally, the ProDiMo model utilizes an enhanced gas/dust ratio of $> 100$ throughout the whole disk, and an even higher ratio within the inner few au . In our case, depleting small dust within the inner tens of 10 au in our model did not appear to rectify the difference between predicted and observed \CO{} $J$=2-1. The gas temperature in the inner 4 au of the ProDiMo model is on the order of 500 K or more, which we do not see in our model, and none of our observations constrain gas temperatures at such small scales. ProDiMo specifically models the inner disk separately from the outer disk, and the temperature within this region is significantly warmer than the outer disk region.

\begin{deluxetable*}{cccc}
\label{tab:upperlevelCO}
\tablecolumns{4}
\tablewidth{0pt}
\tablecaption{CO upper level transitions}
\tablehead{\textbf{Transition} & \textbf{Model Flux} & \textbf{Observed Flux} & \textbf{ Model in }\\
& \multicolumn{2}{c}{Integrated Intensity [1 $\times$ 10$^{-17}$ W m$^{-2}$]}& \textbf{Observed Range?}}
\startdata
5-4 & 0.504 & 1.04 $\pm$ 0.4 & 2$\sigma$ \\
6-5 & 0.677 & 0.74 $\pm$ 0.29 & 1$\sigma$ \\
7-6 & 0.807 & 0.9 $\pm$ 0.3 & 1$\sigma$ \\
8-7 & 0.847 & 1.24 $\pm$ 0.55 & 1$\sigma$ \\
9-8 & 0.766 & 0.91 $\pm$ 0.4 & 1$\sigma$ \\
10-9 & 0.628 & 1.17 $\pm$ 0.35 & 2$\sigma$ \\
11-10 & 0.566 & 1.13 $\pm$ 0.35 & 2$\sigma$ \\
12-11 & 0.586 & 1.17 $\pm$ 0.35 & 2$\sigma$ \\
13-12 & 0.606 & 1.52 $\pm$ 0.4 & 3$\sigma$ \\
14-13 & 0.626 & < 1.6  & True \\
15-14 & 0.648 & 1.03 $\pm$ 0.5 & 1$\sigma$ \\
16-15 & 0.668 & < 1.3  & True \\
17-16 & 0.680 & 0.75 $\pm$ 0.25 & 1$\sigma$ \\
18-17 & 0.682 & < 0.9 & True \\
19-18 & 0.671 & < 0.9 & True \\
20-19 & 0.65 & < 0.9 & True \\
21-20 & 0.621 & < 0.9 & True \\
22-21 & 0.587 & < 0.9 & True \\
23-22 & 0.549 & < 0.9 & True \\
\enddata
\tablecomments{Observed flux values are from \citet{Fedele16} and references therein. Quoted errors are the 1$\sigma$ noise measured in the continuum nearby each line. The right-most column shows whether the modeled flux agrees within 1, 2, or 3$\sigma$ of the observation, or `True' if the modeled flux is below an upper limit.}
\end{deluxetable*}

%\begin{deluxetable*}{cccc}
%\label{tab:test}
%\tablecolumns{4}
%\tablewidth{0pt}
%\tablecaption{CO upper level transitions}
%\tablehead{\textbf{Transition} & \textbf{Observed Flux} & \textbf{Model Flux} & \textbf{ Model in }\\
%& \multicolumn{2}{c}{Integrated Intensity [1 $\times$ 10$^{-17}$ W m$^{-2}$]}& \textbf{Observed Range?}}
%\startdata
%5-4 & 1.04$\pm$0.4 & 0.476 & 2$\sigma$ \\
%6-5 & 0.74$\pm$0.29 & 0.629 & 1$\sigma$ \\
%7-6 & 0.9$\pm$0.3 & 0.732 & 1$\sigma$ \\
%8-7 & 1.24$\pm$0.55 & 0.74 & 1$\sigma$ \\
%9-8 & 0.91$\pm$0.4 & 0.618 & 1$\sigma$ \\
%10-9 & 1.17$\pm$0.35 & 0.434 & 3$\sigma$ \\
%11-10 & 1.13$\pm$0.35 & 0.254 & 3$\sigma$ \\
%12-11 & 1.17$\pm$0.35 & 0.202 & 3$\sigma$ \\
%13-12 & 1.52$\pm$0.4 & 0.16 & False \\
%14-13 & 1.6 $>$ & 0.127 & True \\
%15-14 & 1.03$\pm$0.5 & 0.0968 & 2$\sigma$ \\
%16-15 & 1.3$>$& 0.068 & True \\
%17-16 & 0.75$\pm$0.25 & 0.0434 & 3$\sigma$ \\
%18-17 & 0.9$>$& 0.0248 & True \\
%19-18 & 0.9$>$ & 0.0124 & True \\
%20-19 & 0.9$>$ & 0.00519 & True \\
%21-20 & 0.9$>$& 0.00225 & True \\
%22-21 & 0.9$>$ & 0.00213 & True \\
%23-22 & 0.9$>$ & 0.00241 & True \\
%\enddata
%\end{deluxetable*}

\bibliography{sample63}{}
\bibliographystyle{aasjournal}

\newpage
%\bibliography{sample63}{}
%\bibliographystyle{aasjournal}

\end{document}